\UseRawInputEncoding
\documentclass[journal, 12pt, peerreview, draftclsnofoot, onecolumn]{IEEEtran}
\usepackage{graphicx}
\usepackage{subfigure}
\usepackage{amsfonts}
\usepackage{amssymb}
\usepackage{amsmath}
\usepackage{bbding}
\usepackage{cite}
\usepackage{mathrsfs}
\usepackage{array}
\usepackage{bbm}
\usepackage{bm}
\usepackage{pifont}
\usepackage{epstopdf}

\begin{document}

\title{GMM-based Symbol Error Rate Prediction for Multicarrier Systems with Impulsive Noise Suppression}

\author{N. Rozic,~\IEEEmembership{Senior Member,~IEEE}, P. Banelli, ~\IEEEmembership{Member,~IEEE}, D. Begusic,~\IEEEmembership{Senior Member,~IEEE} and J. Radic,~\IEEEmembership{Member,~IEEE}\\}

\maketitle

\begin{abstract}
Theoretical analysis of orthogonal frequency division multiplexing (OFDM) systems equipped at the receiver by a non-linear impulsive noise suppressor is a challenging topic in communication systems. Indeed, although an exact closed-form expression for the output signal-to-noise ratio (SNR) of such OFDM systems is available for widely used impulsive noise models, theoretical analysis of the associated symbol error rate (SER) is still open. So far, the analytical SER expressions  available in the literature approximate the time-domain impulsive noise, as Gaussian distributed in the discrete frequency domain. 
Conversely, this work presents an accurate analysis of the distortion noise at the nonlinearity output exploiting a Gaussian mixture model (GMM). By using GMMs we unified the approach of SER prediction for unmitigated systems, as well as for the mitigated ones, equipped by non-linear impulsive noise suppressors, including 
blanking, clipping, clipping-blanking and attenuating processors. 
Closed form expressions for the SER are derived both for non-fading and frequency-selective Rayleigh and Rician fading channels affected by impulsive noise which is represented by GMMs, thus including Bernoulli-Gaussian (BG), Middleton Class-A, as well as (approximated) alpha-stable noise. Theoretic SER performance are compared with simulations, showing very good agreement for all the impulsive noise scenarios and the non-linear suppressors.



\end{abstract}

\section{Introduction}
\label{sec:Intro}

Multicarrier  modulations (MCM) are widely employed in most of the wired and wireless communication systems. 
Among the several reasons for their widespread use, one is certainly their higher resistance to impulsive noise (ImpN) with respect to single carrier modulations (SCM) \cite{Ghosh_1996}.
The evaluation of the output SNR performance, and the associated SER, is an active research area for communication engineers, especially in non-ideal conditions, such those faced in ImpN environments. 
For instance, \cite{Price_1958,Minkoff_1985,Banelli_2000} investigated the effects of non-linear distortions in OFDM systems. Furthermore, the effect of impulsive noise has been investigated on the system performance of coded SCM \cite{Miyamoto_1995}, uncoded SCM and MCM \cite{Ghosh_1996}, as well as for coded MCM in power line communications (PLC) \cite{Ma_2005,Amirshahi_2006}, including Bernoulli-Gaussian (BG), Middleton Class-A \cite{Middleton_1977,Middleton_1999}, and alpha-stable noise models  \cite{Shao_1993,Middleton_1999}. Some fundamental results in the analysis of their performance limits can be found in \cite{Seo_1989,Haring_2004,Zhidkov_2006} where authors derived the achievable pairwise error probability (PEP) bounds for OFDM systems affected by impulsive noise represented by a GMM. More recently, performance analysis for MCM/OFDM systems equipped with impulsive noise suppressors  either for 4G cellular systems (UMTS-LTE), broadcasting (DVBT), local area networks (WiFi), or PLC, have been derived in \cite{Zhidkov_2006,Zhidkov_2008, Banelli_2015,Darsena_2015,Banelli_2016,Rozic_2017}, respectively. Therein, sample-based non-linear suppressors simply clip, or null, or both clip and null, the complex envelope of the  baseband received signal, when it overpasses some given thresholds. Both the thresholds and the parameters of these simple receivers are typically optimized to maximize the output SNR. More specifically, ImpN
 mitigation relies on non-linear instantaneous transformations of the signal, such as the optimal Bayesian estimator (OBE) \cite{Banelli_2013}, and sub-optimal thresholding based approaches, such as blanking \cite{Vastola_1984,Zhidkov_2006,Banelli_2013, Banelli_2016}, clipping-blanking \cite{Suraweera_2003,Zhidkov_2008}, or optimal Bayesian attenuating/clipping devices \cite{Rozic_2017}. Furthermore, the performance of space-time coded systems over frequency-selective multiple-input multiple-output (MIMO) channels impaired by Middleton Class-A impulsive noise, is evaluated in \cite{Gao_2007}, \cite{Savoia_2013}.

Actually, the SER closed-form expressions available in the literature are typically developed assuming that the distortion noise at the ImpN suppressor output has a Gaussian distribution. However, this approximation is acceptable only when the average number of noise impulses per OFDM/MCM frame is sufficiently high to invoke the central-limit theorem \cite{Zhidkov_2006,Banelli_2015}. To overcome this issue, the authors in \cite{Banelli_2015} proposed a semianalytical SER analysis for instantaneous ImpN removal in frequency-selective fading channels, which turns out to be quite accurate. Similarly, authors in \cite{Rugini_2005} presented a semianalytical approach to evaluate the error probability of OFDM systems subject to carrier frequency offset (CFO) in frequency-selective channels, characterized by Rayleigh or Rician fading. 

This work proposes a unified GMM-based analytical approach for both unmitigated and  mitigated OFDM systems affected by ImpN. For mitigated systems, the residual distortion noise at the output of the non-linear suppressor (NLS) is modeled with a $\mathcal{K}$-GMM, where $\mathcal{K}$ is not directly related with $\mathit{K}$-GMM ImpN in the channel\footnote{in this work, $\mathit{K}$ denotes the number of ImpN components in the actual channel, while $\mathcal{K}$ rather denotes the number of GMM components used to approximate the distortion noise at the ImpN suppressor output.}. The proposed framework addresses the SER analysis of the most interesting non-linear impulsive noise suppressors such as the OBE, the Genie-aided detector (GAD), as well as  the thresholding-based Bayesian attenuating suppressor (BAS), thus including as particular cases simple estimators, such as blanking, clipping, and clipping-blanking. Specifically, we derive SER analytical expressions for flat and frequency-selective channels, both in non-fading and Rayleigh/Rician fading environments affected by $K$-GMM impulsive noises such as BG, Class-A, and GMM-approximated alpha-stable noises. The theoretical SER performance has been compared with the results obtained by simulation, showing very good agreement for all the realistic impulsive noise scenarios. 
Although the paper focuses on OFDM systems, the theoretical framework and conclusions can be easily extended to any other MCM system.

The first essential and innovative contribution of this manuscript is the novel GMM-based unified approach for prediction of the SER in multicarrier (OFDM) systems, affected by arbitrarily distributed ImpN, with or without ImpN mitigation. To the best of our knowledge, such generalized framework is new in the scientific literature. Furthermore, in the presence of ImpN mitigation, we propose a GMM approximation of the output distortion noise, which exploits a novel 
component-by-component representation. This way, we derive accurate closed-form analytical expressions for the associated SER, improving the  accurancy w.r.t. conventional approaches based on approximated assumptions of Gaussian distribution for the residual noise or, equivalently, on sufficiently long OFDM blocks to accommodate a sufficiently high number of ImpN events \cite{Zhidkov_2006, Banelli_2016}.

The rest of the paper is organized as follows. Section II briefly summarizes the system and noise models. Section III derives SER performance for frequency-flat non-fading OFDM system affected by K-GMM ImpN, in the absence of any impulse noise suppressor, where AWGN (1-GMM) and 2-GMM  \cite{Ghosh_1996} are two particular cases. Section IV extends the analysis to systems employing non-threshold based suppressors, namely the GAD \cite{Eriksson_1995} and OBE \cite{Banelli_2013}. Then, the SER analysis for threshold-based ImpN suppression is derived in Section V. 
Section VI extends the analysis to frequency-selective channels in the presence of slow Rayleigh/Rician fading, when the system is equipped by ImpN NLSs. Numerical and simulation results are presented in Section VII, including comparison with some SER prediction alternatives, while Section VIII provides some concluding remarks. Appendices A and B are dedicated to detailed proofs and mathematical derivations.

\section{System and Noise Models}
\label{sect2:The Model}

This work focuses on M-QAM multicarrier (OFDM) communications over frequency flat and frequency-selective, possibly fading, channels, impaired by impulsive noise. Thus, in this section we introduce the associated system and noise models.

\subsection{System Model}

\begin{figure*}[h]
  \centering
  \includegraphics[scale=1.2]{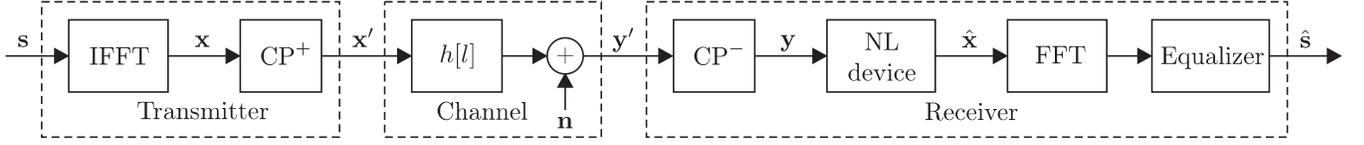}
  \caption{Block scheme of the baseband system model}
  \label{fig:System-Model}
\end{figure*}

Let's define  $\mathbf{s}_q=\left[s_q[0] ,s_q[1] ,\cdots ,s_q[L-1] \right]^{T} $ the \textit{q}th information symbol to be transmitted on $L$ orthogonal subcarriers of an OFDM system, as shown in Fig. \ref{fig:System-Model}. Assuming the data $\left\{s_q\left[l\right]\right\}_{l=0,...,L-1}$ on different subcarriers are independent, with zero mean and same variance $\sigma _{s}^{2} =E\left\{\left|s_q\left[l\right]\right|^{2} \right\}$, the time-domain OFDM symbol $\mathbf{x}_q=\left[x_q\left[0\right],x_q\left[1\right],\cdots ,x_q\left[L-1\right]\right]^{T} $ is obtained by  $\mathbf{x}_q=\mathbf{F}_L^{H}\mathbf{s}_q$, where $\mathbf{F}_L$ is the \textit{L}-point unitary DFT matrix \cite{Wang_2000,Banelli_2014} and $(\cdot)^H$ is the Hermitian operator. The inter-symbol interference is avoided by adding a cyclic prefix (CP) with length $L_h$ to the OFDM block, resulting in $\mathbf{x}_q^{\textrm{cp}}=\left[x_q\left[L-L_{h}\right],\cdots ,x_q\left[L-1\right],x_q\left[0\right],x_q\left[1\right],\cdots ,x_q\left[L-1\right]\right]^{T}$. Then, after D/A conversion and RF modulation, the OFDM block is transmitted through a frequency-selective channel, with discrete-time impulse response $\mathbf{h}_q=\left[h_q\left[0\right],h_q\left[1\right],\ldots ,h_q\left[L_{h}-1\right], 0,\ldots,0 \right]^{T}$. The frequency-flat channel is simply a  special case when $h[0]=1$ and $\left\{h\left[l\right]=0\right\}_{l=1,\cdots ,L-1}$, and consequently \eqref{eq:y_q[n]=Sum} simply reduces to $y\left[l\right]=x\left[l\right]+n[l]$. In this case, the CP$^+$ box in Fig. 1 at the transmitter, as well as the CP$^-$ and "Equalizer" boxes at the receiver, are omitted.
Rayleigh fading is a typical channel model used to mimic worst-case scenario in non-LoS conditions, assuming each $h_{q}[l]$ is distributed as a complex zero-mean Gaussian RV, with stationary variance $\sigma_{l}^{2} =E\left\{\left|h_q\left[l\right]\right|^{2} \right\}$. Conversely, in the presence of Line of Sight (LoS), $E\left\{h_{q}(0)\right\} \ne 0$ makes the statistics of $|h_{q} [0]|$ a Rician one. More general models include the Nakagami-n models \cite{Rice_1948, Nakagami_1960}.

In this paper, we consider Rayleigh and Rice slow (quasi-static) fading, where all the $L_{h}$ active paths are assumed as time-invariant during the transmission of each OFDM block (block-fading model). The extension of the proposed analysis to fast fading where the $L_{h}$ active paths are time-variant within each OFDM symbol (double-selective channels), although quite straightforward, it is out of the scope of this paper.

After CP removal, the received OFDM symbol    $\mathbf{y}_q=[y_q\left[0\right],y_q\left[1\right],\cdots ,y_q\left[L-1\right]]^{T}$  is obtained by a cyclic convolution with the channel, as expressed by
\begin{equation} \label{eq:y_q[n]=Sum}
y_{q} \left[l\right]=h_{q} \left[l\right]\otimes_{L} x_{q} \left[l\right]+n_{q}[l],
\end{equation}

\noindent
which in the frequency domain corresponds to carrier by carrier multiplication of the channel transfer function with the transmitted data, as expressed by
\begin{equation} \label{eq:y_q(f)}
\mathbf{y}_q^{(f)}=\mathbf{F}_L \mathbf{y}_q=\textrm{diag}\left(\mathbf{h}_q^{(f)}\right)\mathbf{s}_q + \mathbf{n}_q^{(f)},
\end{equation}
where $\mathbf{h}_q^{(f)}= \sqrt{L} \mathbf{F}_L \mathbf{h}_q$ and $\mathbf{n}_q^{(f)}=\mathbf{F}_L \mathbf{n}_q$ are the frequency-domain channel response and noise, respectively, during the \textit{q}th OFDM block. Assuming the block-fading channel paths are uncorrelated, the variance of the received signal $y_q[l]$ is given by $\sigma_{y_q}^{2} = \sum_{l=0}^{L-1}\left|h_q\left[l\right]\right| ^{2}\sigma_{x}^{2}+\sigma_{n}^{2}$ where $\sigma_{x}^{2}=E\{\left|x_q\left[l\right]\right|^2\}$ and $\sigma_{n}^{2}=E\{\left|n_q\left[l\right]\right|^2\}$ are the transmitted OFDM signal and receiver noise variances, respectively. 
Zero-forcing (ZF) equalization is applied in the frequency domain, simply dividing  
$\mathbf{y}_q^{(f)}$ in \eqref{eq:y_q(f)} by $\mathbf{h}_q^{(f)}$, leading to
\begin{equation} \label{eq:y_qe(f)}
\mathbf{y}_{q,e}^{(f)}=\mathbf{s}_q + \left(\textrm{diag}\left(\mathbf{h}_q^{(f)}\right)\right)^{-1} \mathbf{n}_q^{(f)}.
\end{equation} 

As shown in Fig. \ref{fig:System-Model}, the instantaneous non-linear suppressor (NLS), which mitigates the noise impulses in the time-domain before DFT processing, spreads in the frequency-domain, over all the OFDM subcarriers, the distortion introduced by the time-domain sparse noise impulses. 

The NLS output, which can be formally expressed by
\begin{equation} \label{eq:x[n]=g(y[n];b[n])}
\hat{x}_q[l]=g(y_q[l];\bm{\pi}_q), \end{equation}

\noindent
depends on the vector $\bm{\pi}_q= \left\{\sigma_n,  \sigma_x,  \Sigma_l|h_q[l]|^2 \right\}$,  which contains the set of parameters that influence the statistical property of the \textit{q}th OFDM block, such as the signal power $\sigma_{y_q}^2$, and the noise power $\sigma_n^2$.

\subsection{Impulsive Noise Models}

This paper focuses on impulsive noise modeled by a $K$-GMM \cite{Kassam_1988}, whose probability density function ($pdf$) $f_{N}(n)$ is expressed by
\begin{equation} \label{eq:f_w-KGMM}
 f_{N} (n)=\sum_{k=0}^{K-1}p_{k} G(n,\sigma _{k}^{2}),
\end{equation}
\noindent
where
$\{p_{k}\}_{k=0,1,\cdots,K-1}$, with $\sum_{k=0}^{K-1}p_{k}=1$,
are the probabilities that an impulsive event with variance $\sigma _{k}^{2}$, is generated according to the \textit{k}th Gaussian $pdf$ $G(n,\sigma _{k}^{2})$. The component $k=0$ represents the "white" noise contribution, appearing with probability $p_{0}$ and variance $\sigma_{0}^{2}$, while the other $K-1$ components represent the impulsive noise, which appears with probability $p_I=\sum_{k=1}^{K-1}p_{k}=1-p_{0}$ and total variance $\sigma_I^2=\sum_{k=1}^{K-1}p_{k}\sigma _{k}^{2}$.

The reason to choose  model \eqref{eq:f_w-KGMM} are twofold. Firstly, $K$-GMM in \eqref{eq:f_w-KGMM} is a mixture of Gaussian functions, thus ensuring analytical tractability and elegance of calculation, and secondly, $K$-GMM can successfully fit any ImpN distribution either exactly \cite{Caire_2014, Hur_2016}, or approximately \cite{Blackard_1993, Sanchez_1999}.

In the following we will focus on widely used ImpN models that can be described by \eqref{eq:f_w-KGMM}, including the Bernoulli-Gaussian (BG) model, the  Middleton's Class-A model, and the symmetric alpha-stable (S$\alpha$S) model.
However, the approach presented in this work can be obviously applied to any other noise that is modelled as $K$-GMM.

\subsubsection{BG Impulsive Noise}

The BG model \cite{Ghosh_1996,  Zhidkov_2006} corresponds to a 2-GMM ImpN, where 
a noise sample is considered either thermal or impulsive with  probabilities $p_0$ and $p_1=1-p_{0}$, respectively. The noise $W\sim {\it G}(w,\sigma_{0}^{2} )$ and the impulse noise $I\sim {\it G}(i,\sigma_{1}^{2})$ are both zero-mean independent Gaussian RVs, resulting in ImpN with a total variance $\sigma_{N}^2=p_0 \sigma_{0}^2 + p_1 \sigma_{1}^2$. AWGN is the even simpler case with $K$=1 and $p_0=1$.

Actually, the 2-GMM can also well approximate Class-A ImpN for $A\ll 1$ and, under proper conditions, can be used as a simple approximative model for the distortion noise at the output of the NL ImpN suppressors, which works noticeably better than the AWGN approximation in \cite{Zhidkov_2008},\cite{Banelli_2015}.

\subsubsection{Class-A Impulsive Noise}

The Middleton's canonical Class-A model \cite{Middleton_1977}, \cite{Middleton_1999} assumes ImpN with $K=\infty$, as a sum of the two independent components $W$ and $I$, where $W$ is a  stationary Gaussian (background) noise with variance $\sigma_W^2$, and $I$ is the impulsive noise with variance $\sigma_I^2$ generated independently by interfering sources (simultaneous, possibly infinite). Only a specific one of the K components could be present in each noise sample with a  probability $p_k$ modeled according to a Poisson distribution, as expressed by
\begin{equation} \label{eq:p_k-Poisson}
p_{k} =e^{-A} A^{k} /k!,
\end{equation}
\noindent
where $A$ is a canonical parameter, also known as the impulsiveness index. 
  
From \eqref{eq:p_k-Poisson}, when $k=0$ it follows $p_0=e^{-A}$, which represents the probability that the channel experiences only the thermic noise $W$. Otherwise, with probability $1-p_0$, the channel experiences an impulsive noise, whose k-$th$ component variance is $\sigma_{k}^2=\sigma_{W}^2 + \sigma_{I}^2 \frac{k}{A}$.

Obviously, a canonical Class-A model is generally compliant with  \eqref{eq:f_w-KGMM}. The "white" noise component is defined by $p_0$ and $\sigma_0^2$, the impulsive component by $p_I=1-p_0$ and $\sigma_I^2=\sum _{k=1}^{\infty }p_{k} \sigma_{k}^{2}/(1-p_0)$, and the total ImpN variance by $\sigma _{n}^{2} =\sum _{k=0}^{\infty }p_{k} \sigma_{k}^{2}=p_0 \sigma _{0}^{2}+p_I \sigma _{I}^{2}$.

For realistic systems, the ${K}$-GMM in \eqref{eq:f_w-KGMM} with $K \ge 10$ is generally a very good approximation of the Class-A model. However, instead of simple truncation of the Poisson distribution as proposed in \cite{Vastola_1984}, \cite{Lin_2013}, we prefer using a more accurate approximation which preserves the first two moments, which is obtained by fitting the approximated $\bar{p}_k$, $\bar{\sigma}_{k}^2$ according to
\begin{equation} \label{eq:bar-p_k)} \bar{p}_k=p_k / \sum _{k=0}^{K-1}p_{k}, \hspace{3pt} \bar{\sigma}_{k}^2=\sigma_{k}^2 / \sum _{k=0}^{K-1}p_{k}\sigma_{k}^2,
\end{equation}
\noindent
which keeps unchanged the signal-to-impulsive noise ratio (SIR) defined as
\begin{equation} \label{19)} \text{SIR}=\frac{\sigma _{x}^{2} }{\sigma_{I}^{2} },  
\end{equation} 
\noindent
where $\sigma _{I}^{2} =\sum _{k=1}^{K } \bar{p}_{k} \bar{\sigma} _{k}^{2}/(1-p_0)  $ represents to the impulsive noise variance.

\subsubsection{$\alpha$-Stable Impulsive Noise}

An $\alpha$-stable random variable models an ImpN with a so-called heavy-tailed (algebraic) distribution. Generally, the $\alpha$-stable \textit{pdf} is defined by a set of  four canonical parameters $(\alpha, \beta, \gamma, \mu)$ where  the characteristic exponent $\alpha$ controls the heaviness of the $pdf$ tails, $\beta$ controls the skewness, while $\gamma=\sigma^{\alpha}$ measures the \textit{pdf} spread around its mean $\mu$.
Unfortunately, an explicit analytical expression for the $\alpha$-stable \textit{pdf} does not exist, except for some specific values of alpha including $\alpha=0.5$ (Cauchy \textit{pdf}) and  $\alpha=2$ (Gaussian \textit{pdf}), which is the \textit{pdf} with minimal thickness. However, due to its generality, $\alpha$-stable distributions have found wide interest in statistics and communications \cite{Shao_1993,Feller_1971}.

Indeed, also an $\alpha$-stable distribution can be analytically defined by an infinite mixture of Gaussians \cite{Hamdan_2004}, that can be actually approximated by a finite scaled Gaussian mixture in practice \cite{Kuruoglu_1998}. An efficient approximation of the standard symmetric $\alpha$-stable S$\alpha$S distribution ($\beta=0, \gamma=1, \mu=0$) by the $K$-GMM presented in \cite{Rozic_2013b} is realized through computing the weighting factors $w_k$ and variances $\sigma_k^2$ by minimizing the relative entropy between the $K$-GMM pdf and the S$\alpha$S pdf in the amplitude range $\pm A_{\rm{N}}$.

\section{SER Analysis for Frequency-flat OFDM Systems Without ImpN Suppression}
\label{sec:SER1}

Assuming that a $K$-GMM impulsive noise affects OFDM symbols in a  frequency-flat channel ($h[0]=1, h[l]=0 \quad \forall l\geq 1$, in Fig.\ref{fig:System-Model}), and the receiver does not perform any kind of ImpN mitigation (NLS block in Fig.\ref{fig:System-Model} is omitted), an exact analytical expression for the SER can be derived as a generalization of the approach in \cite{Ghosh_1996}.

\subsection{SER for \text{K-GMM} ImpN Without Suppression} 

Let's consider a $K$-GMM ImpN with associated probabilities and variances $\left\{p_k, \sigma _{k}^{2} =\gamma _{k} \sigma _{x}^{2} \right\}_{k=1,\cdots ,K-1} $  \footnote{$\gamma_k$ obviously denotes the relative variance $\gamma _{k} = \frac{\sigma _{k}^{2}} {\sigma _{x}^{2}}$.}. Let $K-1$ impulse components affect $l\le L$ samples of an $L$-sized OFDM block, and the rest of $L-l$ samples are affected only by the Gaussian noise component with variance $\sigma _{0}^{2} =\gamma _{0} \sigma _{x}^{2} $, whose probability is $p_{0} =1-\sum _{k=1}^{K-1}p_{k}  $. The probability that $l\le L$ samples are affected by an impulsive noise $I$, possibly composed by at most $K-1$ components, is given by the multinomial distribution \cite{Evans_2000}
\begin{equation} \label{1)} w_{\mathbf{l_{K-1}}}=w\left(l_{1} ,\cdots ,l_{K-1}  \right)=\frac{l!}{l_{1} !\cdots l_{K-1} !} q_{1}^{l_{1} } \cdots q_{K-1}^{l_{K-1} }
\end{equation}
where $\mathbf{\text{l}}_{K-1}=[l_{1} ,\cdots ,l_{K-1}]$, $\sum_{k=1}^{K-1}l_{k}=l$,  and $q_k= \frac{p_k}{1-p_0}$ with $\sum_{k=1}^{K-1}q_{k}=1$.

Let's express by $w(l)=\frac{L!}{(L-l) ! l !} p_{0}^{L-l} p_{I}^{l} $ the binomial distribution that defines the probability to experience an impulsive $l$-tuple event characterized by \eqref{1)} within the L-sized OFDM block. 

Thus, generalizing the approach in \cite{Ghosh_1996} for 2-GMM, the SER for M-QAM square constellation affected by $K$-GMM impulsive noise can be expressed by
\begin{equation} \label{888} p_{s}^{(K-GMM)} =\sum _{l=0}^{L}w(l)\sum _{l_{1} =0}^{l}\sum _{l_{2} =0}^{l-l_{1} }\cdots \sum _{l_{K-1}=0 }^{l-\sum _{m=1}^{K-2}l_{m}  }w_{\mathbf{l_{K-1}}} p_{s;\mathbf{l}_{K}}^{(AWGN)}      
\end{equation} 
\noindent
where 
$p_{s,\mathbf{l}_{K}}^{(AWGN)} =4\frac{\sqrt{M} -1}{\sqrt{M} } B_{\mathbf{l}_{K}} \left(1-\frac{\sqrt{M} -1}{\sqrt{M} } B_{\mathbf{l}_{K}} \right)$,
\noindent 
with $B_{\mathbf{l}_{K}} =Q\left(\sqrt{\frac{3\, \rho_{\mathbf{l}_{K}} }{M-1} } \right)$. Here, $Q\left(x\right)$ denotes the Gaussian Q-function \cite{Simon_2006}, and $\rho_{\mathbf{l}_{K}}$ is the SNR \footnote{note the equality of the output SNR $\rho$ in the time- and the frequency domain.} given by
\begin{equation} \label{888b} \rho_{\mathbf{l}_{K}} =\frac{\sigma _{x}^{2} }{\sigma _{\mathbf{l}_{K}} ^{2} } =\frac{L}{\sum _{k=0}^{K-1}l_{k} \gamma _{k}  }.
\end{equation}  

It should be noted that the computational complexity of \eqref{888} is defined by the number of the $K$-tuple combinations. Even for relatively low $K$, such as $K=5$, the number of combinations is pretty high, and, as discussed in Sect. IIB for Class A ImpN, an approximation with e.g. 4-GMM,  can provide good SER predictions with acceptable complexity, as shown in Sect. VII. In addition, further reduction of the computational complexity can be reached using an $L'<L$ rather than $L$ in \eqref{888}, for instance by considering only those $l$-tuples whose $p_{s,\mathbf{l}_{K}}^{(AWGN)}$ is over $10^{-20}$. This approach is particularly efficient for BG and Class-A ImpN at low $p_I$. On the contrary, for relatively high $p_I$, the simulation results presented in Sect. VII show that accurate SER prediction can be obtained even by using a simple 1-GMM (AWGN) model. However, such conclusions do not hold true for an arbitrarily distributed ImpN, including $S \alpha S$ ImpN.

\subsection{SER Prediction for 1-GMM (AWGN) Approximation}

When $K=1$, expression \eqref{888} reduces to the well known SER formula for M-QAM square constellations in AWGN channels \cite{Proakis_2008}, as expressed by
\begin{equation} \label{ZEqnNum402188} p_{s}^{(AWGN)}(\rho) =4\frac{\sqrt{M} -1}{\sqrt{M} } B_{0}(\rho) \left(1-\frac{\sqrt{M} -1}{\sqrt{M} } B_{0}(\rho) \right), \end{equation}  
\noindent
where $B_{0}(\rho) =Q\left(\sqrt{\frac{3\, \rho }{M-1} } \right)$, and $\rho =\frac{\sigma _{x}^{2} }{\sigma _{n}^{2} } =\frac{1}{\sum _{k=0}^{K-1}p_{k} \gamma _{k} }$ denotes the average SNR.

Under general ImpN conditions, an AWGN model (formally 1-GMM) can actually be employed as a good predictor of the SER, when the total ImpN distribution in the frequency domain can be well approximated by a Gaussian one. As shown in Fig. \ref{fig:SER_CA_origin}, this is the case for Class A ImpN when $A \ge 0.1$. However, for S$\alpha$S ImpN Fig. \ref{fig:SER_SaS_K16_K4} shows that the AWGN model  \eqref{ZEqnNum402188} does not provide accurate SER predictions even when the impulsive noise is highly probable.

\subsection{SER Prediction for 2-GMM Approximation } 

With $K=2$, expression \eqref{888} reduces to
\begin{equation} \label{777} p_{s}^{(2-GMM)}=\sum_{l=0}^{L}w_{l}p_{s}^{(AWGN)}(\rho_l)
\end{equation}

\noindent
where $w_l=\frac{L !} {(L-l) ! l !} p_{0}^{L-l} p_{1}^{l}$ with $\rho_{l} =\frac{\sigma _{x}^{2} }{\sigma _{l}^{2} } =\frac{L}{(L-l) \gamma _{0}+l \gamma _{1}  } $, and $\gamma _{0}=\frac{\sigma_{0}^2}{\sigma_{x}^2}$, $\gamma _{1}=\frac{\sigma_{1}^2}{\sigma_{x}^2}$.

Note that expression \eqref{777}, which is here derived from \eqref{888} for a special case $K=2$, corresponds to the BG ImpN case already considered in \cite{Ghosh_1996}. 

Under some conditions, as discussed in Sect. II, a 2-GMM is an accurate approximation of a $K$-GMM ImpN, and thus can be used as a good SER predictor. Particularly, a Class A ImpN with a relatively low impulse probability, e.g., when the weighting factors for $K>2$ are relatively low, can be safely approximated by a two-component Gaussian mixture.

\section{SER Analysis for Frequency-flat OFDM System with  ImpN Suppression}
\label{sec:SER2}

In order to extend the previous analysis to systems employing non-linear suppression (NLS) of ImpN, as shown in Fig. \ref{fig:System-Model}, it is possible to approximate the non-Gaussian distribution of the output distortion noise with a Gaussian mixture \eqref{eq:f_w-KGMM}. Intuitively, the distortion noise at the output of the NLS may retain some of the features of the impulsive noise  at the suppressor input, because the impulsive noise suppressor can  only mitigate noise impulses, and does not eliminate them completely\footnote{due to the fact that ImpN is a random variable}. Actually, when the input noise is $K$-GMM, and the full noise state information (NSI) is available, the optimal GAD suppressor is linear \cite{Rozic_2017} and, consequently, also the distribution of the distortion noise at the output of the non-linear device is effectively modeled by a $K$-GMM, with the same weighting factors, but different variances. This scenario is considered in subsection B. 

However, in realistic systems, NSI is not fully available and suppressors turn to be non-linear, such as the OBE in \cite{Banelli_2013}. In this case the output distortion noise can not be exactly modeled by a GMM, 
because each single component is no longer Gaussian distributed, and a $\mathcal{K}$-GMM can be used only as an approximation. The use of the OBE is considered in subsection C, and a threshold-based NLS in Section V.

\subsection{SNR at the Output of the NLS}

Considering the non-fading channel, the statistical parameters of the received signal and of the channel noise are fixed over all the OFDM blocks. Then, similarly to \eqref{eq:x[n]=g(y[n];b[n])},\footnote{note that for a non-fading channels we can omit subscript $q$} the output of a memory-free non-linear suppressing device $g\left(y\right)$ is  given by $\hat{x}=g\left(y;\bm{\pi}\right)$. By the Bussgang's theorem \cite{Bussgang_1952,Price_1958,Minkoff_1985}, where $y=x+n$, the output $\hat{x}$ can be expressed as
\begin{equation} \label{eq:x_hat=alpha_x+d}
\hat{x}=g(x+n;\bm{\pi})=\alpha_{\bm{\pi}} x+d,
\end{equation}

\noindent
where $\alpha_{\bm{\pi}} $ is the scaling factor of the information-bearing signal $x$, which depends on the "average" CSI contained in $\bm{\pi}$, and $d$ is the distortion noise.

It follows from \eqref{eq:x_hat=alpha_x+d} that the output distortion noise is given by
\begin{equation} \label{eq:d=x_hat-alpha_x}
d=\hat{x}-\alpha_{\bm{\pi}} x. \end{equation}

For any arbitrary non-linear  device, with real or complex valued input, the average output SNR $\rho$  can be computed by \cite{Banelli_2016}
\begin{equation} \label{ZEqnNum813541} \rho =\frac{E\left\{\left|\alpha x\right|^{2} \right\}}{E\left\{\left|d\right|^{2} \right\}} =\frac{E\left\{\left|\alpha x\right|^{2} \right\}}{E\left\{\left|\hat{x}-\alpha x\right|^{2} \right\}} =\frac{\alpha_{\bm{\pi}} ^{2} \sigma _{x}^{2} }{\sigma _{d \bm{\pi}}^{2} }  \end{equation} 

\noindent
where $\sigma _{d\bm{\pi}}^{2}$ is the variance of the output distortion noise in \eqref{eq:x_hat=alpha_x+d}. 

Considering a $K$-GMM ImpN and the SER in \eqref{888} for the associated SNR \eqref{888b}, we can use the same formula in the presence of ImpN mitigation, by expressing \eqref{ZEqnNum813541}, similarly to \eqref{888b},  as
\begin{equation} \label{888c} \rho_{\mathbf{l}_{K}} =\frac{\alpha_{\pi}^{2} \sigma _{x}^{2} }{\sigma _{\mathbf{l}_{K}} ^{2} } =\frac{L\alpha_{\pi}^{2}}{\sum _{k=0}^{K-1}l_{k} \gamma _{k}  }.
\end{equation}

\subsection{SER Prediction for GAD}

If the full NSI is available to the receiver, the genie-aided suppressor \cite{Eriksson_1995} can optimally mitigate the received signal samples according to the actual noise variance. The MMSE optimal genie-aided Bayesian estimator knows which noise component $k\in \left\{0,K-1\right\}$ is active and simply computes the output by $\hat{x}_{\text{GAD}}=\beta_{k}y_k$,  where $\beta_{k}=\frac{1} {1+\gamma_{k}}$ \cite{Banelli_2013}. 

Obviously, the overall output noise is a nonlinearly attenuated version of the input noise, while the useful output signal is  attenuated by the factor  $\alpha_{\text{GAD}}=\sum_{k=0}^{K-1}{p_{k}}\beta_{k}$ \cite{Rozic_2017} by means of the Bussgang' theorem. Given $\hat{x}$ and $\alpha_{\text{GAD}}$, the output SNR is  \cite{Rozic_2017}
\begin{equation} \label{eq:SNR_GAD}
\rho^{\text{GAD}} =\frac{\left[\sum_{k=0}^{K-1}p_{k} \beta_{k}\right]^{2} }{\sum_{k=0}^{K-1}p_{k} (1+\gamma_{k}) \beta_{k}^{2} -\left[\sum_{k=0}^{K-1}p_{k} \beta_{k}\right]^{2} }.
\end{equation}

Given \eqref{eq:SNR_GAD}, the SER performance can be confidently computed either by \eqref{777} for BG ImpN ($K=2$), or by \eqref{888} for $K\ge 3$.

Note that the SER granted by the GAD, which exploits full NSI at the receiver (e.g., which noise component is active at any sample), is the best achievable SER.  

\subsection{SER Prediction for OBE}

When NSI is limited, i.e., the receiver does not know which noise component is active at any time, the optimal Bayesian estimator (OBE) is obtained by a non-linear signal attenuation, whose optimal parameters are exactly defined by simple closed-form analytical expressions \cite{Banelli_2016,Rozic_2017}. However, the output performance, including SNR and SER, can not be expressed in a closed-form because the output noise distribution is not Gaussian, as well as each one of its components. 

As proved in \cite{Rozic_2017}, the output SNR, and consequently $\alpha_{\text{OBE}}$, can be closely approximated with the output SNR, and $\alpha_{\text{BAS}}$, characterizing the BAS output, when the number of thresholds $M_T$ is high enough ($M_T \ge 10$). Therefore, in the analysis we can exploit the following approximation 
\begin{equation} \label{eq:items_BAS}
\alpha_{\text{OBE}} \approx \alpha_{\text{BAS}}=2\sigma_x^2 \sum _{m=0}^{M}\beta_{m} \sum_{k=0}^{K-1}p_{k} (a_{m,k} -a_{m+1,k}),
\end{equation}

\noindent
where $a_{m,k} =\left[1+A_{m}^{2}/2(1+\gamma_k)\sigma_{x}^{2} \right]e^{-{A_{m}^{2}}/2(1+\gamma_k)\sigma_{x}^{2}}$, as detailed in \cite{Rozic_2017}.
 
Given $\rho_{\mathbf{l}_K}$ \eqref{888c} and $\alpha_{\text{OBE}}$ \eqref{eq:items_BAS}, the SER for the optimal Bayesian estimator directly follows from \eqref{888}.

Note that when only limited NSI is available to the receiver, the SER performance granted by the OBE, is the best acievable one  \cite{Banelli_2013}.

\section{SER Analysis for Frequency-flat OFDM Systems with Threshold-based Suppressors}
\label{sec:SER BAS}

This section analyses the SER performance for the ideal, frquency-flat channel scenario ($h[0]=1, h[l]=0, \forall l\geq 1$), when the NLS block in Fig.\ref{fig:System-Model} performs ImpN mitigation by using a (sub)optimal single thresholding, based on its (sub)optimal detection of the channel state (e.g. "white noise" or "impulse noise").

Common thresholding suppression techniques include blanking (nulling) \cite{Zhidkov_2006}, clipping-blanking  \cite{Zhidkov_2008}, Bayesian attenuating or Bayesian clipping \cite{Rozic_2017}, and their combinations with multiple thresholds. Such threshold-based suppressors may be preferred to OBE due to their relatively low complexity and high efficacy. 
We consider the simplest scenario where the receiver employs a single threshold to perform a binary hypothesis test on the noise state, based on the received signal envelope $|y|$. Specifically, the receiver assumes the state "signal+AWGN" when $|y| \le A_T$, and conversely "signal+impulse noise" when $|y| > A_T$. As detailed in the following subsections, the distribution of the output distortion noise for this single-threshold suppressor can be generally approximated by $\mathcal{K}$-GMM.

\subsection{Mixture Model for the Distribution of the Distortion Noise }

Since the $K$-GMM ImpN variable $N$ includes "white" noise $W$ and impulsive noise $I$ components, both  mitigated by the single-threshold non-linear suppressor, the noise at the suppressor output is composed of four possible components, each belonging to one of the following events 
\begin{equation} \label{ZEqnNum571474} \begin{array}{l} {\; \left. \begin{array}{l} {y_{1} =x+w} \\ {y_{2} =x+i} \end{array}\right\}{\rm when}\; \left|y\right|\le A_{T} } \\ {\; \left. \begin{array}{l} {y_{3} =x+w} \\ {y_{4} =x+i} \end{array}\right\}{\rm when}\; \left|y\right|>A_{T} }, \end{array}
\end{equation} 
\noindent
where \textit{y} is the complex base-band received signal sample at the input of the non-linear impulsive noise suppressor, and $A_T$ is the single ImpN detecting threshold.

Consequently, an adequate model is a four-component scaled mixture, whose $pdf$ is expressed by
\begin{equation} \label{ZEqnNum522403} f_{D} \left(d\right)=\sum _{k=1}^{4}w_{k} f_{D_{k} } \left(d\right),
\end{equation} 
\noindent
where  $\left\{f_{D_{k}(d) } \right\}_{k=1,\cdots ,4}$ and $\left\{w_{k} \right\}_{k=1,\cdots ,4} $  are the distortion noise $pdf$s  and the associated weighting factors of the different components in \eqref{ZEqnNum571474}, respectively.

\begin{figure}[h]
  \centering
  \includegraphics[width=9.0cm, height=6.8cm]{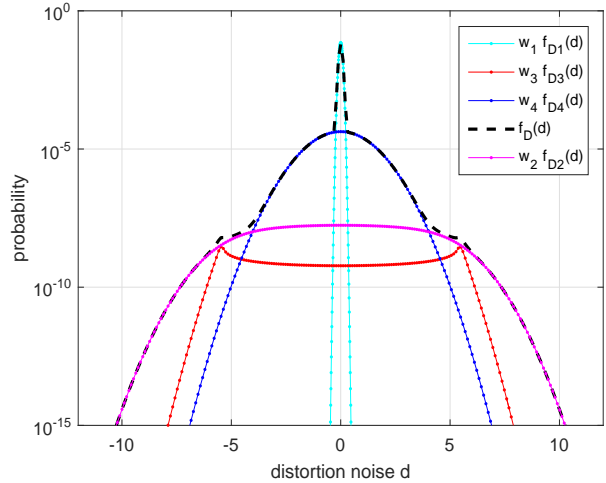}  \caption{Distortion noise distributions $f_{D_k}(d); k=1,..,4$ for BG ImpN at SIR=-10 dB, $p_1=0.01$, SNR=25 dB}
  \label{fig:CL_14_Fig4h}
\end{figure}

\begin{figure}[h]
  \centering
  \includegraphics[width=9.0cm, height=6.8cm]{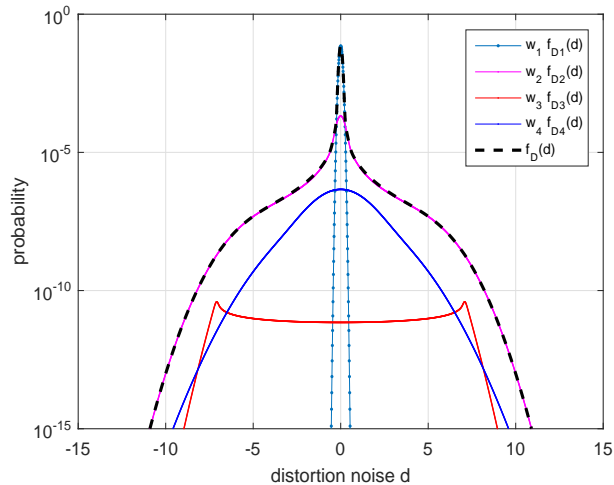}  
  \caption{Distortion noise distributions $f_{D_k}(d); k=1,..,4$ for S$\alpha$S ImpN at $\alpha=1.2$, SIR=-10 dB, $p_1=0.01$, SNR=25 dB}
  \label{fig:pdf_SaS_Fig1}
\end{figure}

Given the threshold $A_{T} $, the weighting factors $w_{k}$ in \eqref{ZEqnNum522403} can be easily computed from $w_1=$ Prob$\left\{|y_1| \leqslant A_T|N=W \right\}$ $\times$ Prob$\left\{N=W\right\}$, etc., leading to
\begin{equation} \label{ZEqnNum197659} \begin{array}{l} {w_{1} =p_{0} \int _{0}^{A_{T} }p_{\left|Y_{0} \right|}  \left(y\right)dy=p_{0} \left(1-c_{0} \right)} \\ {w_{2} =p_{1} \int _{0}^{A_{T} }p_{\left|Y_{I} \right|}  \left(y\right)dy=p_{1} \left(1-c_{1} \right)} \\ {w_{3} =p_{0} \int _{A_{T} }^{\infty }p_{\left|Y_{0} \right|}  \left(y\right)dy=p_{0} c_{0} } \\ {w_{4} =p_{1} \int _{A_{T} }^{\infty }p_{\left|Y_{I} \right|}  \left(y\right)dy=p_{1} c_{1} }, \end{array}
\end{equation} 

\noindent
where $c_{0} =\exp \left(-\frac{A_{T}^{2} }{\sigma _{y_{0} }^{2} } \right)$, $c_{1} =\exp \left(-\frac{A_{T}^{2} }{\sigma _{y_{I} }^{2} } \right)$, and $\sigma _{y_{0} }^{2} =2\left(\sigma _{x}^{2} +\sigma _{0}^{2} \right)$, $\sigma _{y_{I} }^{2} =2\left(\sigma _{x}^{2} +\sigma _{I}^{2} \right)$.

The $pdf$s $\left\{f_{D_{k}(d) } \right\}_{k=1,\cdots ,4}$, as well as their variances $\left\{\sigma _{D_k}^{2} \right\}_{k=1,\cdots ,4}$ are derived analytically in Appendix A. 

Figs. \ref{fig:CL_14_Fig4h} and \ref{fig:pdf_SaS_Fig1} show the shape of the analytical results of $f_{D_k}(d)$, when either BG or S$\alpha$S ImpNs affect the OFDM receiver input. Note that the component distributions in  Figs. \ref{fig:CL_14_Fig4h} and \ref{fig:pdf_SaS_Fig1} are scaled with their weighting factor $w_k$ in \eqref{ZEqnNum197659}, in order to show their impact on the total $f_D(d)$ in \eqref{ZEqnNum522403}.

\subsection{Gaussian Mixture Model for $f_{D}(d)$}

The proposed approximation of the $f_{D}(d)$ by $\mathcal{K}$-GMM leverages on a 
component-by-component Gaussian fitting, as detailed in Appendix B. The closed-form expressions for the $\mathcal{K}$-GMM parameters, e.g., the dimension $\mathcal{K}$, the  variances $\left\{\sigma _{G_k}^{2} \right\}_{k=1,\cdots ,\mathcal{K}}$, and the  scaling factors $\left\{w_{G_k} \right\}_{k=1,\cdots ,\mathcal{K}} $, allow to compute the SER for any ImpN scenario, by 
\eqref{888} and \eqref{888b}, where $\gamma_{k}=\frac{\sigma_{G_k}^2}{\sigma_{x}^2}$ and $w_{k}=w_{G_k}$, for $k=1,\cdots ,\mathcal{K}$.
The results of the $\mathcal{K}$-GMM fitting of $f_{D}(d)$ for any ImpN scenario show that an order $\mathcal{K}\in \left\{2,3,4\right\}$ is typically enough, as verified in Section VII, which shows quite accurate SER estimates.

It is important to stress that the GMM approximation of the time domain distribution $f_{D}(d)$ clearly shows that the 1-GMM (AWGN) can not be considered as a good fitting model for any ImpN scenario.  
Thus, as demonstrated in Sect.VII, and confirmed by results available in the literature  \cite{Zhidkov_2006}, \cite{Banelli_2015}, conventional expressions for $M$-QAM error probability in AWGN channels \cite{Proakis_2008} can be effectively used for OFDM systems only under certain ImpN scenarios. That is, only when the output distortion noise in the frequency domain can be safely approximated as a Gaussian RV \cite{Zhidkov_2006,Banelli_2015}, which is approximately true when  
the average number of noise impulses per OFDM block is relatively high, and the noise is BG or Class-A distributed. Conversely, when an alpha-stable ImpN affects the OFDM channel, even with relatively high average number of impulses per OFDM block, such AWGN approximation does not generally hold true.

\section{SER Performance of OFDM in Frequency-selective Fading Channels Affected by ImpN}
\label{sec:FSC}

The SER performance analysis for fading systems affected by ImpN can be developed based on the results obtained for the non-fading ones. For frequency-selective block-fading channels the SER in (10) for OFDM systems affected by $K$-GMM ImpN, can be still used for each fading realization and each subcarrier, by replacing  
$p_{s}^{(AWGN)}(\rho_l)$ with a proper $p_{s}^{(fading)}(\rho_l)$ accounting for the specific fading channel model.

Indeed, following the system model presented in Fig. 1, and assuming that $h[l], \forall l$ are independent Gaussian RVs, all the $h^{(f)}[l]$ of the frequency domain channel response will be equally distributed.\footnote{this is simplifying assumption where all subcarriers are active (no guard-bands). In general, we can compute $SER_l$ per each subcarrier and then take an average.} 

Consequently, assuming the same signal constellation over all the sub-carriers, the SNR $\rho_{q}[l]$ on the $q$th block will be the same for any carrier $l=1,\dots,L$. Thus, to simplify notation, in the following we omit the subcarrier index $l$ and we use the $\rho_{q}$ to denote the SNR in the $q$th block for any subcarrier.

The slow (quasi-static) frequency-selective block fading channel, will make the signal power $\sigma_{x_q}^2$ fluctuating from an OFDM block to another, according to the \textit{q}th realization $\{h_q\left[l\right]\}_{l=0,\ldots,L-1}$ of the channel. Thus, if all the \textit{L} subcarriers transmit independent data $s_q\left[l\right]$\footnote{digital systems regularly use scrambling before modulation \cite{Proakis_2008}}, the time-domain signal power $\sigma_{x}^2$ is stationary (after CP removal), and the conditional $pdf$ $f_{x_q|\mathbf{h}_q}\left(x,\mathbf{h}_q\right)$ of the useful component $x_q\left[l\right]$ for a given channel $\mathbf{h}_q$ is a Gaussian function $G(x,0,\sigma_{x_q}^2)$, where  $\sigma_{x_q}^2=E\{|x_q\left[l\right]|^2\}=g_q \sigma_{x}^2$, and $g_q=\sum_{l=0}^{L-1}{|h_q[l]|^2}$ is the power attenuation induced by the channel during the $q$-th OFDM block. Assuming that $g_q$ varies in a block-fading fashion, while the noise variance $\sigma_{n}^2$ is fixed, the received signal power is expressed by $\sigma_{y_q}^2=g_q\sigma_{x}^2+\sigma_{n}^2$.

Consequently, in order to compute the average SER for slow (block) fading channels equipped with NLS, it is necessary to develop the expression for the conditional distribution of the signal-to-noise ratio $\rho_q=\sigma _{x_q}^{2}/\sigma _{n}^{2}$. Let $f(\rho_q;\bar{\rho}_{q})$ denote the required $pdf$ of $\rho_q$ for the $q$th block when the average signal-to-noise ratio is $\bar{\rho}_{q}$, and $p_{s} (\rho _{q})$ denote the conditional SER, then the unconditional SER follows from
\begin{equation} \label{151} p_{s}^{(fading)}=\int _{0}^{\infty }f\left(\rho _{q} ;\bar{\rho }_{q} \right) p^{(AWGN)}_{s} (\rho _{q}) d\rho _{q},
\end{equation} 
\noindent
under the assumption that the overall noise effect can be modelled as Gaussian.

Specifically, considering a single threshold ImpN suppression, the optimal threshold $A_{T_q}$, as well as the associated attenuating parameters $\beta_{0_q}$ and $\beta_{1_q}$, should adapt their values to maximize the system performance \cite{Rozic_2017}.

\subsection{Rayleigh Fading}

Rayleigh fading represents a multipath channel, constituted by a number of simultaneous time-aligned weak reflected components without any direct line-of-sight (LOS) path.
 
Let $\bar{\rho }_{q}$ denote the average SNR during the $q$-th OFDM block at the NLS output\footnote{Note that, the average signal-to-noise $\bar{\rho_q }$ can be simply computed from \eqref{ZEqnNum813541}}. Then, the instantaneous value $\rho _{q} $ for the $q$th OFDM block is exponentially distributed as \cite{Alouini_1999, Yen_2007}
\begin{equation} \label{17)} f\left(\rho _{q} ;\bar{\rho }_{q} \right)=\frac{1}{\bar{\rho }_{q} } e^{-\rho _{q} /\bar{\rho }_{q} } ,\quad \bar{\rho }_{q} \ge 0.
\end{equation} 

Actually, in Rayleigh fading scenarios, as shown in \cite{Banelli_2015}, the total output noise can be well approximated by a Gaussian $pdf$. 
Thus, given \eqref{17)}, the unconditional SER $p_{s}^{Ray}(\bar{\rho}_{q})$ can be computed from \eqref{151} by 
\begin{equation} \label{ZEqnNum919824} p_{s}^{Ray}(\bar{\rho}_{q})=\int _{0}^{\infty }\frac{1}{\bar{\rho }_{q} } e^{-\rho _{q} /\bar{\rho }_{q} } p_{s}^{(AWGN)} \left(\rho _{q} \right)d\rho_{q},
\end{equation}
\noindent
where $p^{(AWGN)}(\rho_{q})$ is expressed by \eqref{ZEqnNum402188} \cite{Yen_2007}  
for square $M$-QAM constellation.

By exploiting in \eqref{ZEqnNum919824} the alternative $Q$-function formulation  \cite{Craig_1991}
\begin{equation} \label{20)} Q(x)=\frac{1}{\pi } \int _{0}^{\pi /2}e^{-\frac{x^{2} }{2\sin ^{2} \phi } } d\phi ,\quad x\ge 0,  \end{equation} 
\noindent
the average SER has the following closed-form expression 
\begin{equation} \label{ZEqnNum229768} \begin{array}{l} {p_{s}^{(Ray)}(\bar{\rho}_{q})=1-\frac{1}{M} -2\frac{\sqrt{M} -1}{M} \frac{a}{\sqrt{1+a^{2} } } } \\ {\quad \quad -\frac{4}{\pi } \left(\frac{\sqrt{M} -1}{\sqrt{M} } \right)^{2} \frac{a}{\sqrt{1+a^{2} } } \tan ^{-1} \frac{a}{\sqrt{1+a^{2} } } }, \end{array} \end{equation} 
\noindent
where $a=\sqrt{\frac{3\bar{\rho_{q} }}{2\left(M-1\right)} } $, and $\bar{\rho}_{q}$.

Thus, exploiting the good Gaussian approximation for the noise in Rayleigh fading scenario, the noticeably simpler expression \eqref{ZEqnNum229768} can be effectively used, rather than (10), to get a good SER approximation. This is due to the fact that the fading-compensated AWGN noise dominates the suppressed ImpN, as it can be easily verified by comparing the distortion noise powers with that ones obtained in flat channels.

\subsection{Rician Fading}

Differently to Rayleigh fading, in Rician fading the received signal includes a direct LoS component, together with a number of Gaussian, typically weaker, reflected components. Considering a slow (quasi-static) frequency-selective block fading
channel, the SNR per $q$th 
block is distributed according to a noncentral chi-square $pdf$, as expressed by \cite{Alouini_1999,Yen_2007}
\begin{equation} \label{ZEqnNum134658} f\left(\rho _{q} ;\bar{\rho }_{q} ;K_{r} \right)=\frac{K_{1} e^{-K_{r} } }{\bar{\rho }_{q} } e^{-K_{1} \frac{{\rho_q}}{\bar{\rho }_{q} }} I_{0} \left(\sqrt{4K_{r} K_{1} \frac{\rho _{q} }{\bar{\rho }_{q} }  } \right),
\end{equation}

\noindent
where $K_1=(1+K_r)$, and $K_{r} $ is the ratio of the LoS path with the scattered powers. When the channel power is normalized to 1, $K_{r} =\bar{h}_{0}^{2} /(1-\bar{h}_{0}^{2} )$, where $\bar{h}_{0} $ is the average LoS path gain. 

As it happens with Rayleigh fading channels, if we consider SER results for $K_{r} \le 10$ and typical ImpN parameters, the fading interference dominates the distortion noise caused by the (sub)optimal NLS and consequently, the total output noise can be considered approximately Gaussian. Thus, rather than using \eqref{888}, also in this case it is possible to derive a much simpler expression 
to closely approximate the SER. 

Specifically, when $K_{r} \le 10$, substituting \eqref{151} in \eqref{ZEqnNum134658}, the unconditional SER can be approximated by
\begin{equation} \label{160)} p_{s}^{(Rice)}(\bar{\rho_{q}} ,K_r) =\int _{0}^{\infty }f\left(\rho_{q} ;\bar{\rho_{q}} ;K_r \right)p_{s}^{(AWGN)}(\rho_q) d\rho _{q},  \end{equation}
\noindent
which can be conveniently expressed in the following equivalent form \cite{Shayesteh_1995, Alouini_1999} 
\begin{equation} \label{28)} p_{s}^{(Rice)}(\bar{\rho_{q}} ,K_r)=\frac{e^{-K_{r} } }{\pi } \left[aI_{1}(\bar{\rho_{q}} ,K_r) +bI_{2}(\bar{\rho_{q}} ,K_r) \right], \end{equation} 
\noindent
where the parameters $\left(a,b,c\right)$ depend on the modulation and the signal constellation, $I_{1}(\rho,K_r) $ and $I_{2}(\rho,K_r) $ are given by
\begin{equation} \label{ZEqnNum130504} \begin{array}{l} {I_{1}(\rho,K_r) =\int _{0}^{\pi /2}h_{\rho,K_r}\left(\theta \right)e^{K_{r} h_{\rho,K_r}\left(\theta \right)} d\theta  \ } \\ {I_{2}(\rho,K_r) =\int _{0}^{\pi /4}h_{\rho,K_r}\left(\theta \right)e^{K_{r} h_{\rho,K_r}\left(\theta \right)} d\theta,  \ } \end{array} \end{equation} 
\noindent
and $h_{\rho,K_r}\left(\theta \right)$ is a function of the SNR and Rician factor, as expressed by
\begin{equation} \label{ZEqnNum299614} h_{\rho,K_r}\left(\theta \right)=\frac{\left(K_{r} +1\right)\cos ^{2} \theta }{\left(K_{r} +1\right)\cos ^{2} \theta +\bar{\rho }c^{2} } \ . \end{equation} 

Specifically, for 4-QAM and 16-QAM the parameters $\left(a,b,c\right)$ are expressed by \cite{Shayesteh_1995}

\hspace{10pt}4-QAM: $a=b=1$, $c=\sin \left(\pi /4\right)$,\\
\indent\indent
16-QAM: $a=3/4,\; b=9/4$, $c=\sqrt{0.1} $.

Unfortunately, closed-form analytical solutions of the  integrals in \eqref{ZEqnNum130504} are not available. Thus in order to obtain a closed form expression, we need to resort to an approximate solution by using the middle Riemann sum \cite{Hughes_2005}, obtaining
\begin{equation} \label{ZEqnNum856574} \begin{array}{l} {I_{1}(\rho,K_r) = \beta \sum _{m=0}^{M_{0}-1}h_{\rho,K_r}\left(m \beta\right)e^{K_{r} h_{\rho,K_r}\left(m \beta \right)}} \\ {I_{2}(\rho,K_r) =\beta \sum _{m=0}^{M_{0}/2-1}h_{\rho,K_r}\left(m \beta \right)e^{K_{r} h_{\rho,K_r}\left(m \beta \right)},} \end{array} \end{equation}  

\noindent
where $M_{0}$ is the number of samples\footnote{as experienced with a number of simulations with different parameters, good results can be obtained by only $M_0 =11$ with middle point $\rho = \bar{\rho}$.} used in the numerical method, and $\beta=\frac{\pi }{2M_{0}}$.

Thus, the average SER in Rician fading channels \eqref{28)} is approximated by
\begin{equation} \label{ZEqnNum114816} \begin{array}{l} {p_{s}^{(Rice)}(\bar{\rho_{q}} ,K_r)=} \\ {\frac{e^{-K_{r} } }{\pi } \left[a \beta \sum _{m=0}^{M_{0}-1} h_{\bar{\rho_{q}} ,K_r}\left(m \beta \right)e^{K_{r} h_{\bar{\rho_{q}} ,K_r \left(m \beta \right)}} \right.}  \\ { +\left. b \beta \sum _{m=0}^{M_{\theta}/2-1}h_{\bar{\rho_{q}} ,K_r}\left(m \beta \right)e^{K_{r} h_{\bar{\rho_{q}} ,K_r}\left(m \beta \right)}  \right]}, \end{array} \end{equation} 
\noindent
where the average SNR $\bar{\rho}$ in \eqref{ZEqnNum299614} is still given by \eqref{ZEqnNum813541}.

However, when $K_r >10$, the SER results computed from \eqref{ZEqnNum114816} are not acceptable and, similarly to \eqref{888}, it is necessary to resort to more complex expression making use of \eqref{ZEqnNum114816} for $p_{s,\mathbf{l}_{K}}^{(Rice)}$, rather than $p_{s,\mathbf{l}_{K}}^{(AWGN)}$, leading to 
\begin{equation} \label{9999} p_{s}^{(Rice_{K-GMM})} =\sum _{l=0}^{L}w(l)\sum _{l_{1} =0}^{l}\sum _{l_{2} =0}^{l-l_{1} }\cdots \sum _{l_{K-1}=0 }^{l-\sum _{m=1}^{K-2}l_{m}  }w_{\mathbf{l_{K-1}}} p_{s;\mathbf{l}_{K}}^{(Rice)}.      
\end{equation} 
\noindent

\section{Numerical and Simulation Results}
\label{sect7:Simulations}

Numerical and simulation results presented in this section consider the SER performance of OFDM with 4-QAM and 16-QAM constellations employing $L$=256 subcarriers  over nonfading and slow-fading frequency-selective Rayleigh/Rician fading channels, affected by BG, Class-A, or general $K$-GMM, including S$\alpha$S ImpN.

The SER performance are analyzed for a wide range of parameters of practical interest, including the SIR, the impulse probability $p_I$, as well as the "white" SNR defined by the signal to "white" noise ratio $\sigma_x^2/\sigma_w^2$.
The presented results include SER performance of unmitigated systems (Sect. III), mitigated nonfading systems with GAD and OBE (Sect. IV), single-threshold mitigated nonfading systems (Sect. V) and frequency-selective Rayleigh and Rician fading channels (Sect. VI).

\subsection{SER Results for Unmitigated Non-fading Systems}

\begin{figure}[h]
  \centering
  \includegraphics[width=9.0cm, height=6.8cm]{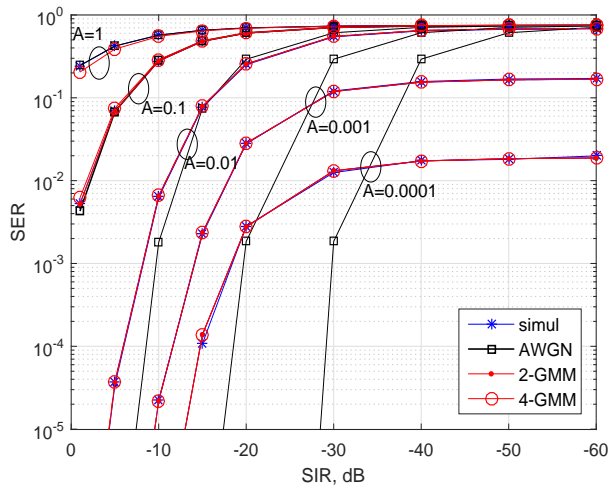}  \caption{SER versus SIR  for 4-QAM OFDM system with $L$=256 subcarriers  affected by Class-A ImpN, restricted to 4 components by using (7), at $A \in \left\{1, 0.1, 0.01, 0.001, 0.0001\right\}$ and SNR = 25 dB.}
  \label{fig:SER_CA_origin}
\end{figure}

Fig.  \ref{fig:SER_CA_origin} shows the SER versus SIR for a 4-QAM OFDM system with $L$=256 subcarriers  affected by Class-A ImpN, accurately approximated by a 4-GMM by using (7). Numerical results computed by \eqref{ZEqnNum402188}, \eqref{777} and \eqref{888} for AWGN, 2-GMM  and 4-GMM, respectively, are compared with simulation results. The SER performance are analyzed for a wide range of signal to impulse noise ratio SIR $\in [-60 \div 0]$ dB and for a wide range of the impulsiveness index $A \in \left\{1, 0.1, 0.01, 0.001, 0.0001\right\}$, at SNR=25dB. Evidently, the results obtained by the 4-GMM \eqref{888}, which is an approximate model of a Class-A ImpN, predict the SER accurately in the whole range of ImpN parameters. 
However, it is also interesting to note that the simple model 2-GMM \eqref{777} is a good SER predictor when $A \le 0.01$. On the other side, the AWGN equivalent model \eqref{ZEqnNum402188} is acceptable only when $A \ge 0.1$.

\begin{figure}[h]
  \centering
  \includegraphics[width=9.0cm, height=6.8cm]{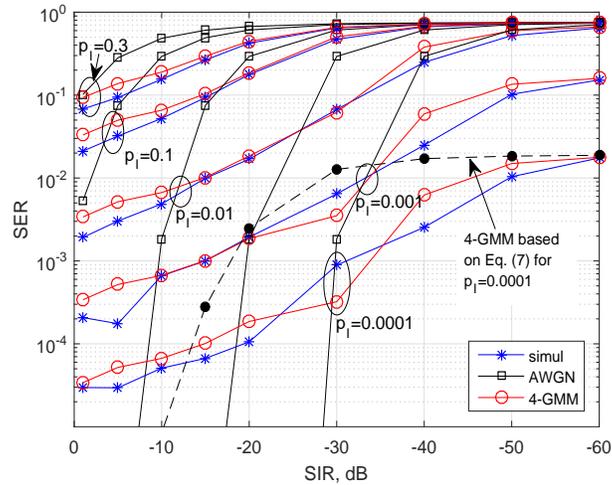}
  \caption{SER versus SIR  for 4-QAM OFDM system with $L$=256 subcarriers  affected by S$\alpha$S ImpN approximated by 4-GMM at $p_1 \in \left\{0.0001,\, 0.001,\, 0.01,\, 0.1,\, 0.3\right\}$ and SNR = 25 dB. Dashed line with black markers shows the SER results obtained by using (7).}
  \label{fig:SER_SaS_K16_K4}
\end{figure}

Fig. \ref{fig:SER_SaS_K16_K4} shows the SER results versus SIR for 4-QAM OFDM systems, with L=256 subcarriers, affected by S$\alpha$S ImpN at $\alpha=1.2$. This S$\alpha$S distribution is accurately approximated by a 16-GMM \cite{Rozic_2013b}. Numerical results, computed by \eqref{ZEqnNum402188} for the AWGN model and by \eqref{888} for the approximate 4-GMM\footnote{Note that the exact ImpN model with K=16 would require a relatively high number of calculations in \eqref{888}.} are compared with the simulation results. The SER performance are analysed in the wide range SIR $\in [-60 \div 0]$ dB and for a wide range of impulsiveness index $A \in \left\{1, 0.1, 0.01, 0.001, 0.0001\right\}$, at SNR=25dB. Similarly to the discussion for Class-A ImpN in Fig. \ref{fig:SER_CA_origin}, expression \eqref{888} with a 4-GMM model, which is an accurate approximation of the exact 16-GMM ImpN, predicts quite well the SER in the whole range of ImpN parameters. Note that the noticeable deviation in the results due to an uneven approximation of the S$\alpha$S distribution by only 4-component GMM at different SIR\footnote{Note that the SER results, obtained by approximative 4-GMM following (7), as shown in Fig. \ref{fig:SER_SaS_K16_K4}, are far to be acceptable, thus a direct 4-GMM approximation of S$\alpha$S ImpN is performed \cite{Rozic_2013b}.}.

\subsection{SER Results for Optimally Mitigated non-fading Systems: GAD and OBE Suppressors}

\begin{figure}[h]
  \centering
   \includegraphics[width=9.0cm, height=6.8cm]{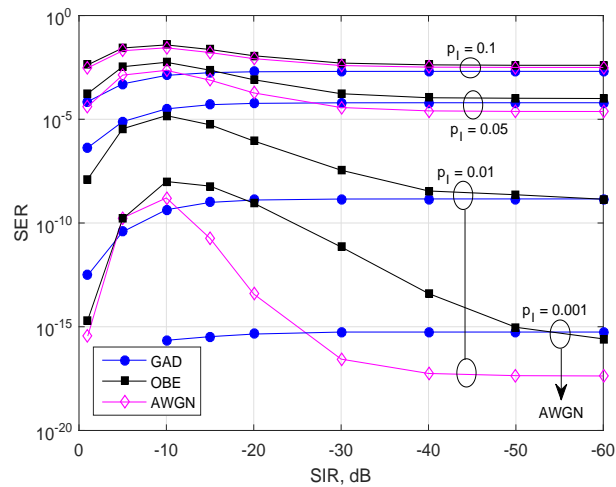}  \caption{The comparison of OBE and GAD suppressors; SER versus SIR at SNR = 25 dB and $p_1 \in \left\{0.001,\, 0.01,\, 0.05,\, 0.1\right\}$ for 4-QAM OFDM with 256 subcarriers affected by BG ImpN.}
  \label{fig:SER_BG_OBE}
\end{figure}

Fig. \ref{fig:SER_BG_OBE} compares numerical results for the SER obtained by OBE and GAD ImpN suppressors. Both of them rely on the optimal Bayesian estimation, however OBE exploits a limited NSI, while the GAD  exploits the full NSI. 
The BG ImpN is considered at SNR=25 dB and at impulse noise probabilities $p_{I} \in \left\{0.1,{\rm \; }0.05,{\rm \; }0.01,{\rm \; }0.001\right\}$ for SIR $\in [-60 \div 0]$ dB. In practice, the SER results for the OBE are obtained approximately by computing the SER for the ten-threshold BAS (see \cite{Rozic_2017}), while SER results for the GAD are computed directly from \eqref{888}, as explained in Sect. IVB. The results for the OBE represent the maximal system performance for a realistic scenario, when the impulse noise positions are not known to the receiver, while SER results for the GAD are the overall best system performance in a genie-aided scenario, which assumes that the impulse positions are exactly known to the receiver. The gain of exploiting such a full NSI is evident, and interestingly, the gain is the highest for those ImpN parameters of highest practical interest in real systems.

\subsection{SER Results for Threshold-based Mitigated Systems}

\begin{figure}[h]
  \centering
  \includegraphics[width=9.0cm, height=6.8cm]{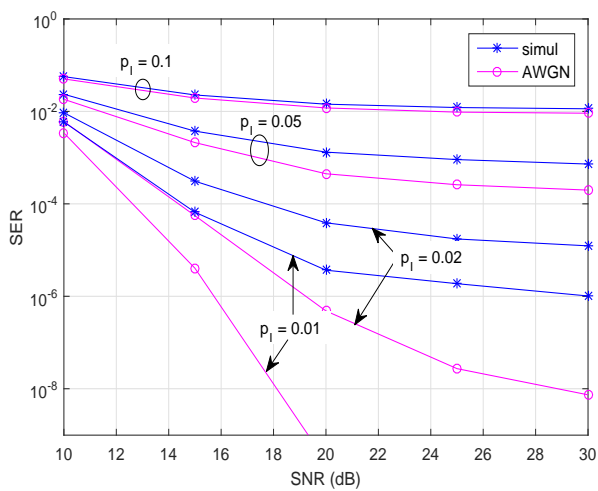}
  \caption{SER versus SNR for SIR=-20 dB and $p_1 \in \left\{0.01,\, 0.02,\, 0.05,\, 0.1\right\}$ for 4-QAM OFDM with 256 subcarriers affected by BG ImpN.}
  \label{fig:SNR_FSC_Fig1b}
\end{figure}

Fig. \ref{fig:SNR_FSC_Fig1b} compares simulation results with the SER obtained by the AWGN model \eqref{ZEqnNum402188},  for 4-QAM OFDM systems with L=256 subcarriers, affected by a BG ImpN, whose ``white'' noise SNR $\in [10 \div 30]$ dB, an impulse component that produces SIR=-20 dB, and impulsive probability $p_1 \in \left\{0.01,\, 0.02,\, 0.05,\, 0.1\right\}$. Evidently, the AWGN equivalent model approximates very well the simulation results only for $p_I \ge 0.1$\footnote{results for $p_I > 0.1$ are not shown, but anticipated from the results up to $p_I = 0.1$.} 
 Conversely, when $p_I<0.1$, such an approximated  SER is acceptable only when SNR$<$10dB.

\begin{figure}[h]
  \centering
   \includegraphics[width=9.0cm, height=6.8cm]{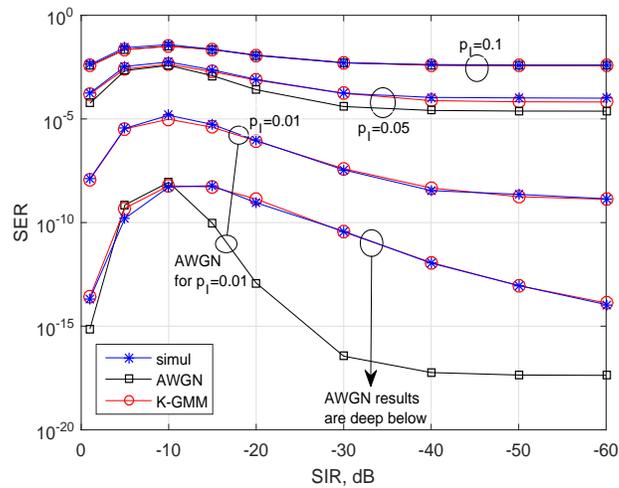} 
   \caption{SER versus SIR at SNR = 25 dB and $p_1 \in \left\{0.001,\, 0.01,\, 0.05,\, 0.1\right\}$ for 4-QAM OFDM with 256 subcarriers affected by BG ImpN.}
  \label{fig:SER_BG_M4}
\end{figure}

Fig. \ref{fig:SER_BG_M4} compares the SER results obtained by AWGN \eqref{ZEqnNum402188}, and $\mathcal{K}$-GMM models \eqref{888} with those obtained through simulation, for a 4-QAM OFDM system equipped with an (sub)optimal single-threshold NLS \cite{Rozic_2017}, and L=256 subcarriers. The ImpN is a BG with a ``white'' noise component at SNR=25 dB and SIR$\in [-60 \div 0]$ dB, with probability $p_{I} \in\left\{0.001,{\rm \; }0.01,{\rm \; }0.05,{\rm \; }0.1\right\}$. Evidently, the $\mathcal{K}$-GMM model \eqref{888} approximates the simulation results very well for every set of ImpN parameters\footnote{Note that all parameters of $\mathcal{K}$-GMM are computed as detailed in Appendix B}. However, for relatively high impulse probability ($p_{I} \geq  0.05$), the results obtained by the AWGN model are acceptable as well. Conversely, for relatively low impulse probability ($p_{I} < 0.05$) the SER predictions obtained by the AWGN model, are not accurate and become quite unacceptable for very low impulse probabilities ($p_{I} \le 0.01$). This is expected since for $L$=256 the average number of impulses per frame for $p_{I} <0.01$ is really low (e.g., 2 or 3 on average), and the distribution of the output distortion noise in the frequency domain is far from a Gaussian one. 

\begin{figure}[h]
  \centering
  \includegraphics[width=9.0cm, height=6.8cm]{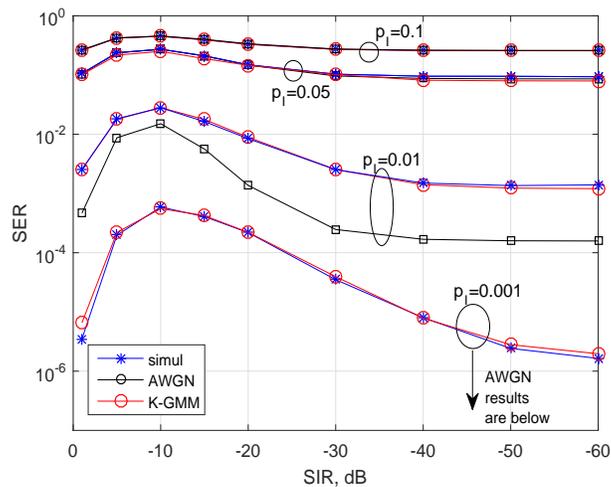}
\caption{SER versus SIR at SNR = 25 dB  for 16-QAM OFDM with 256 subcarriers $p_1 \in \left\{0.001,\, 0.01,\, 0.05,\, 0.1\right\}$ with BG ImpN.}
  \label{fig:SER_BG_M16}
\end{figure}

Fig. \ref{fig:SER_BG_M16}, similarly to Fig. \ref{fig:SER_BG_M4},  compares the SER results obtained by  simulation and by \eqref{ZEqnNum402188} for AWGN  and by \eqref{888} for $\mathcal{K}$-GMM when the OFDM system uses square 16-QAM constellation and $L=256$ subcarriers. The impulsive noise is also BG with the same parameters used in Fig. \ref{fig:SER_BG_M4}, and the overall discussion about its results can be confirmed for 16-QAM OFDM system.

\begin{figure}[h]
  \centering
  \includegraphics[width=9.0cm, height=6.8cm]{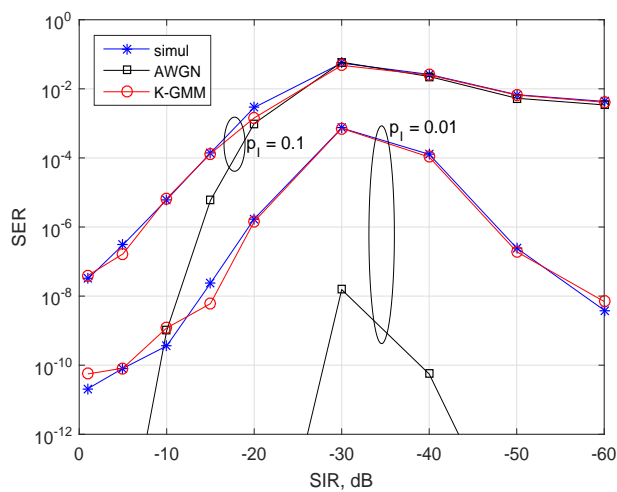}
  \caption{SER versus SIR  for 4-QAM OFDM with 256 subcarriers at SNR = 25 dB and $p_I \in\left\{0.1,\, 0.01\right\}$ with S$\alpha$S ImpN at $\alpha=1.2$.}
  \label{fig:SER_SaS_Fig2}
\end{figure}

Fig. \ref{fig:SER_SaS_Fig2} shows the SER results for the symmetric alpha stable (S$\alpha$S) ImpN, with $\alpha=1.2$ and impulse noise probabilities $p_{I} \in \left\{0.1,{\rm \; }0.01\right\}$. Evidently, equation \eqref{888} for the $\mathcal{K}$-GMM model approximates simulated SER results very well, for every set of impulsive noise parameters. In this case, differently from BG and Class-A cases, the AWGN-based equation \eqref{ZEqnNum402188} does not provide good SER predictions, in particular neither for $p_I=0.1$ or for relatively high SIR (i.e., SIR$\ge -20$ dB). This somewhat unexpected outcome can be explained since, for a 16-GMM fitted to the S$\alpha$S pdf, the  contribution of each component to the total SER is significant. Consequently, there is no component whose contribution to the total SER can be neglected. Under such condition, even for relatively high average number of impulses per frame, and relatively high SIR, each component among the $K-1$ impulsive ones, noticeably contributes to the total SER\footnote{These results point out to the so-called weakness of the central limit theorem}. 
Consequently, the noise distribution in the frequency domain can not be approximated by a Gaussian pdf.

\subsection{SER Results for Frequency-selective Fading Mitigated Systems}

\begin{figure}[h]
  \centering
  \includegraphics[width=9.0cm, height=6.8cm]{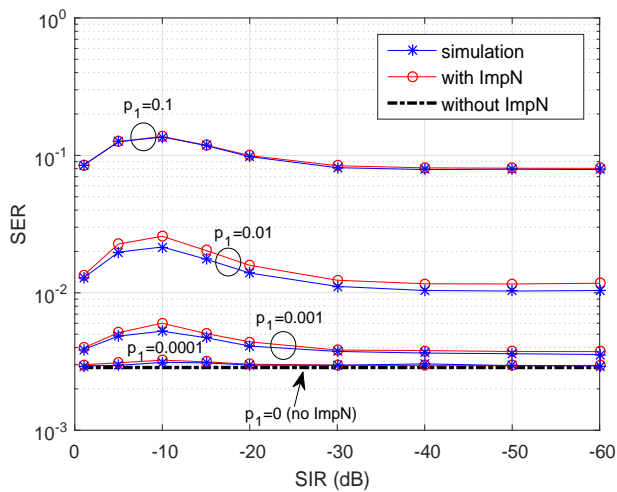}
  \caption{SER versus SIR for SNR=25dB and $p_1 \in \left\{0.0,\, 0.0001,\, 0.001,\, 0.01,\, 0.1\right\}$ for OFDM with 256 subcarriers over slow Rayleigh fading channel affected by BG ImpN.}
  \label{fig:SER_FSC_Fig1}
\end{figure}

Fig. \ref{fig:SER_FSC_Fig1} shows SER versus SIR for SNR=25dB and impulse noise probabilities $p_1 \in \left\{0.0,\,0.0001,\, 0.001,\, 0.01,\, 0.1\right\}$ for slow Rayleigh fading channels,  equipped by (sub)optimal thresholding-based ImpN mitigation. The considered system is a 4-QAM OFDM with $L$=256 subcarriers, affected by BG ImpN. The comparison of numerical results obtained by \eqref{ZEqnNum229768} with simulation results shows good agreement for every SIR.
 
\begin{figure}[h]
  \centering
  \includegraphics[width=9.0cm, height=6.8cm]{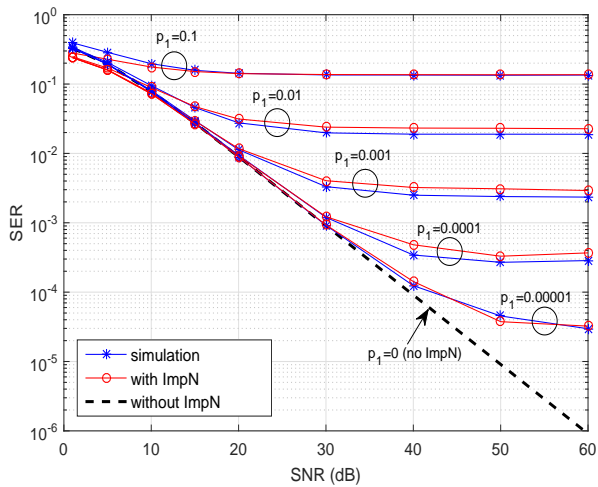}
  \caption{SER versus SNR for OFDM with 256 subcarriers over slow Rayleigh fading channel affected by BG ImpN at SIR=-10dB and $p_1 \in \left\{0.0,\,0.00001,\, 0.0001,\, 0.001,\, 0.01,\, 0.1\right\}$. }
  \label{fig:SER_FSC_Fig2}
\end{figure}

Fig. \ref{fig:SER_FSC_Fig2} shows SER results versus SNR spanning from 0dB to 60dB for a fixed SIR=-10dB, and $p_1 \in \left\{0.0,\,0.00001,\, 0.0001,\, 0.001,\, 0.01,\, 0.1\right\}$. It is evident that the previous comments can be generalized, leading to conclude that also in the case of Rayleigh fading channels, the relatively simple equation \eqref{ZEqnNum229768}, which assumes Gaussian noise in the frequency domain, offers accurate SER predictions, even though the channel is actually affected by ImpN. This result is explained by the fact that, in the time domain, the output Gaussian noise dominantly contributes to the distribution of the distortion noise at the output of an (sub)optimal ImpN suppressor.

\begin{figure}[h]
  \centering
  \includegraphics[width=9.0cm, height=6.8cm]{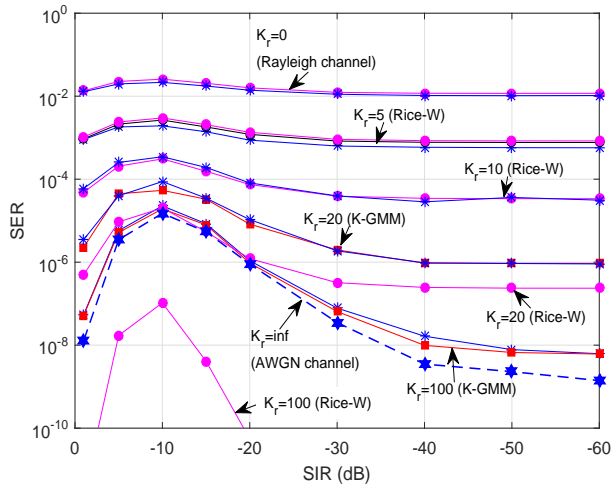}
  \caption{SER versus SIR for SNR=25dB and $p_1=0.01$ for OFDM with 256 subcarriers over Rician fading channel at $K_r\in \left\{0,5,10,100,\infty \right\}$ affected by BG ImpN.}
  \label{fig:SER_Rician_Fig1}
\end{figure}

Fig. \ref{fig:SER_Rician_Fig1} shows the SER versus SIR a for 4-QAM OFDM system with $L$=256 subcarriers equipped, by the Bayesian attenuating ImpN suppressor \cite{Rozic_2017}, at $p_1=0.01$ and SNR=25dB, over the Rician slow fading channel with $K_r \in \left\{0,5,10,100,\infty\right\}$. It is useful to remind that Rician factor $K_r=0$ corresponds to Rayleigh fading, and $K_r=\infty$ corresponds to the AWGN (nonfading) channel. Specifically, Fig. \ref{fig:SER_Rician_Fig1} compares the SER computed by \eqref{ZEqnNum114816} (named Rice-W) and \eqref{9999} (named Rice-$K$-GMM)\footnote{due to simplicity,  here we use $K$ instead of $\mathcal{K}$} with the simulation results, and let us to conclude that for $K_{r} \leq 10$ the simpler Rice-W model works quite well, while for $K_{r}>10$ the Rice-W model underestimates the SER. Conversely, the Rice-$K$-GMM model \eqref{9999} works quite well every time. Obviously, for higher Rician factors $K_r$, the Rice-$K$-GMM model noticeably outperforms in accuracy the Rice-W model, and when $K_r$ approaches infinity it approaches the performance shown in Fig. \ref{fig:SER_FSC_Fig2}, which correspond to the non-fading channel.

\section{Conclusion}
\label{sec:conclusion}

We presented a unified GMM-based analytical approach for SER performance evaluation of OFDM systems affected by  impulsive noise in nonfading and frequency selective Rayleigh/Rice fading channels. The analysis includes  both unmitigated and mitigated systems affected by ImpN modeled with a $K$-GMM, such as BG (2-GMM),  Class-A, as well as impulsive noise with alpha-stable distributions.

For unmitigated non-fading systems it has been shown that the SER can be exactly predicted for any $K$-GMM ImpN. However, at relatively high $K$ (e.g. $K \geq 5$), due to a high number of combinations in the multinomial distribution, the  computation complexity becomes highly time consuming. In such cases we have shown that, an accurate $\mathcal{K}$-GMM ($\mathcal{K} < K$) approximation of the noise distribution, based on first two moments matching, leads to accurate SER estimations.

For mitigated systems, the distortion noise at the output of the NLS is generally non-Gaussian and it can be also  effectively approximated by a $\mathcal{K}$-GMM. Specifically, for threshold-based mitigated systems, the NLS output distortion noise can be safely approximated with at most $\mathcal{K}=4$ Gaussian components. 
The proposed unified approach is based on the $\mathcal{K}$-GMM approximation of the time domain distortion noise distribution $f_{D}(d)$, by performing a novel component-by-component Gaussian fitting method presented in Appendix B. Based on the developed closed-form expressions which exploit the mixture parameters of the fitted $\mathcal{K}$-GMM, the SER can be simply computed for any ImpN scenario, exploiting the  expressions derived for unmitigated systems in Sect. III. The use of appropriate values of $\mathcal{K}$ for different systems and different scenarios guarantees the minimization of the computation complexity. 

Although, we have shown results for BG, Class-A, and alpha-stable ImpNs with the Bayesian non-linear ImpN suppressors, the developed closed-form SER expressions can also be used for other ImpNs and other NLSs, including thresholding-based NLS, such as blanking, clipping, and their combination, as well as the multi-thresholded NLS. The  approach presented in this work can be extended to nonsymetric $pdf$ $f_{D}(d)$ of the distortion noise, will be the subject of our future work.

\appendices

\label{app:appendixA}

\section{Closed-form Expression for Distortion Noise Distribution with Single-threshold Suppressor}

Given the Bussgang's theorem \eqref{eq:x_hat=alpha_x+d}, the distortion noise can be calculated from \eqref{eq:d=x_hat-alpha_x}. 

\begin{figure}[h]
  \centering
    \includegraphics[width=5.0cm]{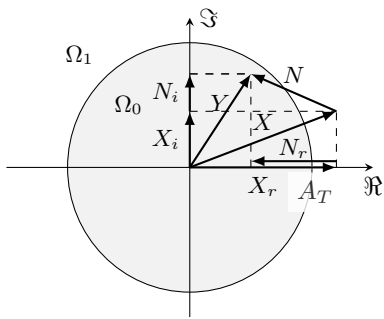}  \caption{Complex-valued signals and thresholding}
  \label{fig:Y_X_N}
\end{figure}

Let the impulsive noise suppressor perform (sub)optimal Bayesian attenuation of the received samples as $\hat{X}=\beta_{\Omega} Y$ \cite{Rozic_2017}. With $Y=X+N$, the time-domain distortion noise $D$ in \eqref{eq:d=x_hat-alpha_x} can be expressed by
\begin{equation} \label{20} D=\left(\beta_{\Omega} -\alpha \right)X+\beta_{\Omega} N.
\end{equation} 
\noindent
where $X$ and $N$, and consequently $Y$, are complex-valued variables, and  $\beta_{\Omega}$ denotes the attenuation factor performed in the signal amplitude range $\Omega$ (see Fig. \ref{fig:Y_X_N}).

Assuming independent $X$ and $N$, the distortion noise distribution $f_{D_{\Omega}} \left(d\right)$ can be calculated by
\begin{equation} \label{ZEqnNum140124} f_{D_{\Omega}} \left(d\right)=f_{\left(\beta_{\Omega} -\alpha \right)X} \left(d\right)*f_{\beta_{\Omega} N} \left(d\right)
\end{equation} 
\noindent
where $(\beta -\alpha) X$ and $\beta N$ are the (sub)optimally attenuated signal and noise, respectively, and $*$ denotes the convolution operator.

Considering the single-threshold ImpN suppressor, the total region $\Omega = \Omega_0 \cup  \Omega_1$ is split into two disjoint subregions $\Omega_0$, $\Omega_1$ where $\Omega_0 \triangleq \left\{y: |y| \leq A_T \right\}$, and $\Omega_1 \triangleq \left\{y: |y| > A_T \right\}$. Then, applying  \eqref{ZEqnNum140124} to each subregion, we write
\begin{equation} \label{22} \begin{array}{l} {f_{D_{\Omega_0}} \left(d \right)=f_{\left(\beta_{\Omega_0} -\alpha \right)X_{\Omega_0}} \left(d \right)* f_{\beta_{\Omega_0} N_{\Omega_0}} \left(d\right)}
 \\ {f_{D_{\Omega_1}} \left(d \right)=f_{\left(\beta_{\Omega_1} -\alpha \right)X_{\Omega_1}} \left(d \right)* f_{\beta_{\Omega_1} N_{\Omega_1}} \left(d \right) }, \end{array} \end{equation}
\noindent
where $X_{\Omega_0}$ and $N_{\Omega_0}$ denote the signal and noise RVs below the threshold $A_{T} $, and $X_{\Omega_1}$ and $N_{\Omega_1}$ denote the signal and noise RVs above the threshold $A_{T}$, respectively. 

Evidently, the distribution of the distortion noise \eqref{22} at the output of the threshold-based ImpN suppressor is expressed by the $pdf$ $f_{(\beta -\alpha) X_{\Omega}} (x)$ and  $f_{\beta N_{\Omega}}(n)$ of the attenuated signal and noise, respectively, evaluated below and above the threshold. 
In the following we firstly develop the closed-form expressions for the distributions of the non-attenuated variables $X_{\Omega}$ and $N_{\Omega}$, and then take into account the attenuating factors in the convolution \eqref{22}  and compute the $pdf$ $f_{D_{\Omega_0}}(d)$, $f_{D_{\Omega_1}}(d)$ of the distortion noise.

Due to thresholding of the received signal amplitude $\left|y\right|=\sqrt{y_{r}^{2} +y_{i}^{2} } $, the detected signal component $X_{r} $ is  related to both the received signal components $Y_{r} $ and $Y_{i} $. Thus, we write
\begin{equation} \label{ZEqnNum702995} \begin{array}{l} {f_{X_{\Omega _{0} } } \left(x\right)=f_{X_{r} } \left(x_{r} \right)\left|_{|y|\le A_{T} } \right. } \\ {\quad \quad =\frac{1}{B} {\displaystyle \iint \nolimits _{\Omega _{0} }} f_{X_{r} Y_{r} Y_{i} } \left(x_{r} ,y_{r} ,y_{i} \right)dy_{r} dy_{i},  } \end{array} \end{equation} 
\noindent
where $B=P\left(\left|y\right|\le A_{T} \right)$.

Assuming that $y_i$ is independent of $x_r$, where,
\[f_{X_{r} Y_{r} Y_{i} } \left(x_{r} ,y_{r} ,y_{i} \right)=f_{X_{r} } \left(x_{r} \right)f_{Y_{r} Y_{i} |X_{r} } \left(y_{r} ,y_{i} |x_{r} \right)\] 
\noindent
and by chain rule, 
\[f_{Y_{r} Y_{i} |X_{r} } \left(y_{r} ,y_{i} |x_{r} \right)=f_{Y_{r}|X_{r} } \left(y_{r} |x_{r} \right)f_{Y_{i} |Y_{r} } \left(y_{i} |y_{r} \right),\] 
\noindent
which substituted in \eqref{ZEqnNum702995} leads to
\begin{equation} \label{12)} \begin{array}{l} {f_{X_{\Omega _{0} } } \left(x\right)=\frac{f_{X_{r} } \left(x_{r} \right)}{B} } \\ {\quad \times {\displaystyle \iint \nolimits _{\Omega _{0}} }f_{Y_{r} |X_{r} } \left(y_{r} |x_{r} \right) f_{Y_{i} |Y_{r} } \left(y_{i} |y_{r} \right.)dy_{i} dy_{r}.  } \end{array} \end{equation} 

The conditional density $f_{Y_{r} |X_{r} } \left(y_{r} |x_{r} \right)$ is obviously imposed by the noise density $f_{N_{r} } \left(n_{r} \right)$, which let us to write
\begin{equation} \label{13)} f_{X_{\Omega _{0} } } \left(x\right)=\frac{f_{X_{r} } \left(x_{r} \right)}{B} \iint \nolimits _{\Omega _{0} }f_{N_{r} } \left(n_{r} \right)f_{Y_{i} |Y_{r} } \left(y_{i} |y_{r} \right)dy_{i} dn_{r}.   \end{equation} 

As illustrated in Fig. 14, it is clear that the integration region for $n_{r} =y_{r} -x_{r} $ is defined by $\pm A_{T} -x_{r} $, and for $y_{i} $ the integration region is defined by $\pm \sqrt{A_{T}^{2} -y_{r}^{2} } $. Thus
\begin{equation} \label{14)} \begin{array}{l} {f_{X_{\Omega _{0} } } \left(x\right)=} \\ {=\frac{f_{X_{r} } \left(x_{r} \right)}{B} {\displaystyle \int} _{-A_{T} -x_{r} }^{A_{T} -x_{r} }f_{N_{r} } \left(n_{r} \right){\displaystyle \int} _{-\sqrt{A_{T}^{2} -y_{r}^{2} } }^{\sqrt{A_{T}^{2} -y_{r}^{2} } } f_{Y_{i} } \left(y_{i} \right) dy_{i} dn_{r}  }. \end{array} \end{equation} 

The $pdf$s $f_{X} \left(x\right)$, $f_{N} \left(n\right)$ and $f_{Y} \left(y\right)$ are Gaussians, while $f_{\left|Y\right|} \left(y\right)$ is Rayleigh, and consequently 
\begin{equation} \label{15)} \int _{-\sqrt{A_{T}^{2} -y^{2} } }^{\sqrt{A_{T}^{2} -y^{2} } }f_{Y} \left(y\right) dy = erf\left(\sqrt{\frac{A_{T}^{2} -y^{2} }{2\sigma _{y}^{2} } } \right), \end{equation}
\noindent
while
\begin{equation} \label{16)} \begin{array}{l} {B=P\left(\left|y\right|\le A_{T} \right)={\displaystyle \int} _{0}^{A_{T} }\frac{\left|y\right|}{\sigma _{y}^{2} }  \exp \left(-\frac{\left|y\right|^{2} }{2\sigma _{y}^{2} } \right)dy}\\ {\quad \quad \quad  \quad \quad \quad \quad \quad \quad \quad \quad =1-\exp \left(-\frac{A_{T}^{2} }{2\sigma _{y}^{2} } \right)}. \end{array} \end{equation}  

Thus exploiting \eqref{15)} and \eqref{16)}, and with $y_r=x_r+n_r$, Eq. \eqref{14)} becomes\footnote{in the following, due to the same distributions of real and imaginary parts, we simply use symbols $X$, $N$ and $Y$ without denoting real or imaginary.}
\begin{equation} \label{ZEqnNum204256} \begin{array}{l} {f_{X_{\Omega_0}} \left(x\right)=\frac{\exp \left(-\frac{x^{2} }{2\sigma _{x}^{2} } \right)}{\left(1-\exp \left(-\frac{A_{T}^{2} }{2\sigma _{y}^{2} } \right)\right) \sqrt{2\pi \sigma _{x}^{2} } }}  \\ {\quad \quad \times {\displaystyle \int \limits_{-A_{T} -x}^{A_{T} -x}}\frac{\exp \left(-\frac{n^{2} }{2\sigma _{n}^{2} } \right)}{\sqrt{2\pi \sigma _{n}^{2} } }  {\kern 1pt}  erf \left(\sqrt{\frac{A_{T}^{2} -\left(x+n\right)^{2} }{2\sigma _{y}^{2}}} \right) dn }. \end{array} \end{equation}

Unfortunately, the integral in \eqref{ZEqnNum204256} can not be analytically solved. Therefore, similarly to the derivation of Eq. \eqref{ZEqnNum114816}, we resort to an accurately approximated numerical solution based on the middle Riemann sum. By using $M_{d} $ equidistant points $d_i$ in the numerical integration within the region $\mp 10\sigma _{y} $, we get
\begin{equation} \label{ZEqnNum338517} \begin{array}{l} {f_{X_{\Omega_0}}(d_i) =\frac{h^{2} }{1-\exp \left(-\frac{A_{T}^{2} }{2\sigma _{y}^{2} } \right)} \frac{\exp \left(-\frac{d_{i}^{2} }{2\sigma _{x}^{2} } \right)}{\sqrt{2\pi \sigma _{x}^{2} } } } \\ {\quad \times {\displaystyle \sum} _{j=-M_{d} -i}^{M_{d} -i}\frac{\exp \left(-\frac{d_{i}^{2} }{2\sigma _{n}^{2} } \right)}{\sqrt{2\pi \sigma _{n}^{2} } }  \, erf\left(\sqrt{\frac{A_{T}^{2} -\left(d_i+d_j\right)^{2} }{2\sigma _{y}^{2} } } \right),} \end{array} \end{equation} 
\noindent
where $M_{d} $ should be relatively high to obtain good results (e.g. $M_{d} =5\cdot 10^{3} $).

 Given \eqref{ZEqnNum204256}, the conditional $pdf$ $f_{X_{\Omega_1}}(x)$, due to a mutual exclusivity of events $|y|\le A_{T} $ and $|y|>A_{T} $, can be simply computed from
\begin{equation} \label{ZEqnNum196507} f_{X_{\Omega_1}}(x)=\frac{f_{X} \left(x\right)-f_{X_{\Omega_0}}(x) P\left(\left|y\right|\le A_{T} \right)}{1-P\left(\left|y\right|\le A_{T} \right)}.  \end{equation} 
\noindent

Given \eqref{ZEqnNum196507}, it is straightforward to compute $f_{X_{\Omega_0}}(d_i) $ at discrete values $x_{i}=d_i$, which is omitted due to lack of space. 

Expressions \eqref{ZEqnNum338517} and \eqref{ZEqnNum196507} calculate accurately $pdf$s below and above the threshold $A_{T} $ for the signal $X$. In the same way, it is straightforward to write the corresponding expressions $f_{N_{\Omega_0}}(d_i) $ and $f_{N_{\Omega_1}}(d_i) $ for the noise $N$ by replacing $x$ and $\sigma _{x}^{2} $ with $n$ and $\sigma _{n}^{2} $ in \eqref{ZEqnNum338517} and \eqref{ZEqnNum196507}. 

Given the closed form expressions for signal and noise $pdf$s 
$f_{X_{\Omega_0}}(d_i) $, $f_{N_{\Omega_0}}(d_i) $ and $f_{X_{\Omega_1}}(d_i) $, $f_{N_{\Omega_1}}(d_i) $, below and above the threshold, respectively, we can compute \eqref{22} by a discrete convolution as 
\begin{equation} \label{ZEqnNum843313} \begin{array}{l} {f_{D_{\Omega_0}} (i)= \sum _{m=-M_{d}}^{M_{d}} f_{\left(\beta_{\Omega_0} -\alpha \right) X_{\Omega_0}} (i-m) f_{\beta_{\Omega_0} N_{\Omega_0}} (m)}
 \\ {f_{D_{\Omega_1}} (i)= \sum _{m=-M_{d}}^{M_{d}} f_{\left(\beta_{\Omega_1} -\alpha \right)X_{\Omega_1}} (i-m) f_{\beta_{\Omega_1} N_{\Omega_1}} (m) .} \end{array} \end{equation}

Finally, assuming that the noise $N$ represents either white noise $W$ with variance $\sigma _{w}^{2} $ and probability $p_{0} $, or impulse noise $I$ with variance $\sigma _{I}^{2} $ and probability $1-p_{0} $, a four-component mixture \eqref{ZEqnNum522403} is completely defined by \eqref{ZEqnNum522403} and \eqref{ZEqnNum843313}, where it holds
\begin{equation} \label{ZEqnNum988512} \begin{array}{l} {f_{D_{1} } \left(d \right)=f_{D_{\Omega_0}} \left(d \right);\; {\rm for}\; N=W} \\ {f_{D_{2} } \left(d \right)=f_{D_{\Omega_0}} \left(d \right);\; {\rm for}\; N=I} \\ {f_{D_{3} } \left(d \right)=f_{D_{\Omega_1}} \left(d \right);\; {\rm for}\; N=W} \\ {f_{D_{4} } \left(d \right)=f_{D_{\Omega_1}} \left(d \right);\; {\rm for}\; N=I.} \end{array} \end{equation} 

Considering $f_{D_{1}}(d)$ as the "white" distortion noise component, the  "impulsive" part $f_{D_{I}} (d)$ is expressed by 
\begin{equation} \label{49)} f_{D_{I}} \left(d\right) = \frac{1}{1-w_{D_1}}  \sum _{k=2}^{K}w_{G_k} f_{G_{k} } \left(d\right). \end{equation}

Given \eqref{ZEqnNum988512}, the variances $\sigma_{D_k}^{2}$ and weighting factors $w_{D_k}$ of the $k$-th component can be accurately calculated from 
\begin{equation} \label{10)} \begin{array}{l} {\sigma_{D_k}^{2}=\sum _{i=-M_{d} /2}^{M_{d} /2} d_{i}^{2}  f_{D_{k} } \left(d_i\right),} \\ {w_{D_k}=w_{k}; \quad k=1,\cdots,4}, \end{array} \end{equation} 
\noindent
where $\{w_{k}\}_{k=1,\cdots,4}$ are given by \eqref{ZEqnNum197659}.

Note that in this Appendix we derived closed-form expressions for $f_{D_{k} } \left(d\right)$, $k=1,\cdots,4$ for 2-GMM ImpN (denoted as BG model). For Class A and generally K-GMM ImpNs, the approach can be easily extended by taking into account all the impulsive components with the corresponding probabilities (weighting factors) $\{p_{k}\}_{k=0,1,\cdots ,K-1}$ and variances $\{\sigma _{k}^{2}\}_{k=0,1,\cdots ,K-1}$. Due to a limited space, the details are omitted.

\label{app:appendixB}

\section{The Proposed GMM Approximation of $f_{D}(d)$} 

As discussed in Section V, the distortion noise $pdf$ $f_{D}(d)$ is generally nonGaussian. However, due to its symmetry, it can be approximated by a mixture of zero-mean Gaussian $pdf$s. Due to the fact that, we are interested in the closed-form expressions for variances and scaling factors of normal random variables, rather than resorting to iterative algorithms, we build the GMM by a novel 
component-by-component approach,

Assuming that $f_{D}(d)$ is a Gaussian mixture, its variance distribution among the potential components $f_{D_k}(d)$, can be calculated from two neighbour samples $d_i$ and $d_{i+1}$ of $\left\{f_{D}(d_i)\right\}_{i=1,\cdots,M_{d}}$ 
by
\begin{equation} \label{ZEqnNum415922} \sigma _{f}^{2}(d_i) =0.5\frac{d_{i+1}^{2} -d_{i}^{2} }{\log f_{D} (d_{i} )-\log f_{D} (d_{i+1} )}.  \end{equation}

\begin{figure}[h]
  \centering
  \includegraphics[width=9.0cm, height=6.8cm]{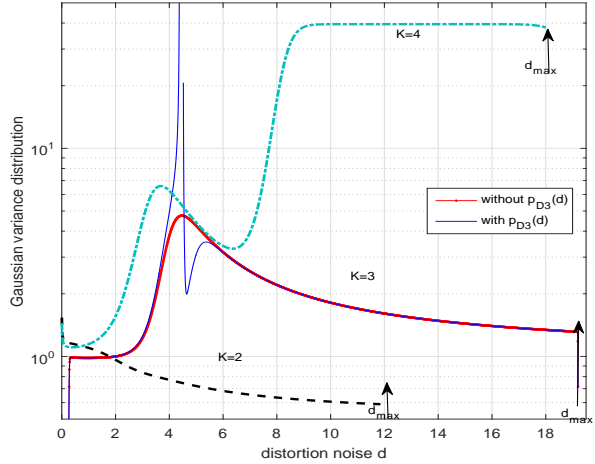}
\caption{Examples of Gaussian variance function $\sigma _{f}^{2}(d_i)$ obtained by \eqref{ZEqnNum415922}; the thin blue line (denoted by $K=3$) shows results obtained with included $f_{D_3}(d)$, and the red tick line when $f_{D_3}(d)$ is excluded. Value $d_{max}$ is the noise amplitude when $f_{D_{I} } (d)$ is lower than the numerical approximation tolerance, e.g. $10^{-15}$.}  
  \label{fig:pD2_BG_Fig1}
\end{figure}

Expectedly, the function $\sigma _{f}^{2}(d_i)$  in \eqref{ZEqnNum415922} is symmetric likewise the $f_{D}(d)$. Fig. \ref{fig:pD2_BG_Fig1} illustrates\footnote{Note that the red line is obtained when $f_{D_3}(d)$ in \eqref{ZEqnNum988512} is omitted in order to eliminate "bursts" in $\sigma _{f}^{2}(d_i)$ (blue line), since the contribution of $f_{D_3}(d)$ to $\sigma _{f}^{2}(d_i)$ can be safely neglected.} the variance function $\left\{\sigma _{f}^{2}(d_i)\right\}_{i=1,\cdots,M_{d}}$, which we exploit to estimate the Gaussian components i.e., their variances $\sigma _{G_k}^{2} $ and scaling weights $w _{G_k}$. 

Given the variances  $\left\{\sigma _{G_k}^{2}\right\}_{k=1,\cdots,\mathcal{K}} $ and the scaling weights $\left\{w _{G_k}\right\}_{k=1,\cdots,\mathcal{K}}$, the total distortion noise distribution $f_{D}(d)$ is approximated by a $\mathcal{K}$-GMM as 
\begin{equation} \label{50)} f_{D} \left(d\right) \approx f_{K-GMM} \left(d\right)=\sum _{k=1}^{\mathcal{K}}w_{G_k} f_{G_{k} } \left(d\right). \end{equation}

\subsection{Component-by-component GMM Approximation of $f_{D}(d)$}

Approximation of the total distortion noise distribution $f_{D}(d)$ \eqref{50)} is performed in a component-by-component way  through at most four steps, as presented in the following.

\noindent
- step 1.

Assuming relatively low probability of impulse noise, the "white" noise distribution $f_{D_1}(d)$ in \eqref{ZEqnNum988512} can be well approximated by Gaussian pdf (see Figs. \ref{fig:CL_14_Fig4h} and \ref{fig:pdf_SaS_Fig1}). Thus, the closed-form expression for the mixture parameters of the first Gaussian component $f_{G_{1}}(d)$ are simply expressed by    
\begin{equation} \label{500} \sigma _{G_1}^{2}=\sigma _{D_1}^{2}, \quad w_{G_{1}}=w_{D_{1}}, \end{equation}
\noindent
where $\sigma _{D_1}^{2}$ and $w_{D_1}$ are given by \eqref{10)} for $k=1$.

Then the "impulsive" distribution $f_{D_{I}}(d)$ expressed by \eqref{49)} is separately approximated by a GMM, as detailed by the following steps. 
As shown in Figs. \ref{fig:CL_14_Fig4h}-\ref{fig:pdf_SaS_Fig1}, for any ImpN scenario, at most three Gaussians are necessary to approximate $f_{D_{I} } (d)$.

\vspace{6 pt}
\noindent
- step 2.

Let's compute the point $d_{G_2}$ given by 
\begin{equation} \label{5002} d_{G_2}=\mathop{\arg }\limits_{d} \left\{f_{D_{I_1} } (d)-c_{0}f_{D} (d)\right\}, \end{equation}
\noindent
where the residual distribution $f_{D_{I_1}}(d)$ given by
\begin{equation}  \label{5003} f_{D_{I_1}}(d)=f_{D} (d)-w_{G_{1}} f_{G_{1} } (d) \end{equation}
\noindent
falls to $c_{0} f_{D}(d)$, when we assume that an additional Gaussian component is needed for proper fitting\footnote{numerical results show that the fitting quality is quite robust to factor $c_{0}=0.9$. Specifically, for GMM fitting to almost GMM $pdf$s, factor $c_0$ should be close to one e.g. 0.99. Generally, an optimal $c_{0}$ would undergo an optimization method,  which however in this work is omitted, because we are interested in closed-form solutions.}. 

Given the Gaussian variance function expressed by \eqref{ZEqnNum415922}, the variance $\sigma _{G_2}^{2}$ of the second Gaussian component $f_{G_{2} } (d)$ can be computed from  
\begin{equation} \label{501} \sigma _{G_2}^{2}= \begin{cases}
 \sigma _{D_I}^{2}   & \quad \text{if } d_{G_2} \geqq d_{max}  \\
 \sigma _{f}^{2}(d_{G_2}) & \quad \text{if } d_{G_2}<d_{max} \end{cases}  \end{equation}\noindent
where $d_{max}$ is the noise amplitude when $f_{D_{I} } (d)$ is lower than the numerical approximation tolerance, e.g., $10^{-15}$ (see  Fig. \ref{fig:pD2_BG_Fig1}).

If \eqref{501} results in $\sigma _{G_2}^{2}=\sigma _{D_I}^{2}$ (case $K$=2 in Fig. \ref{fig:pD2_BG_Fig1}), $f_{D_{I} } (d)$ is approximated by a 1-GMM, and thus the total $f_{D} (d)$ can be well approximated by $\mathcal{K}=2$, resulting in a 2-GMM with the following mixture parameters 
\begin{equation} \label{502} \begin{array}{l} {\sigma _{G_1}^{2}=\sigma _{D_1}^{2}, \quad \sigma _{G_2}^{2}=\sigma _{D}^{2}-w_{D_1} \sigma _{D_1}^{2}}, \\ {w_{G_1}=w_{D_1},\quad w_{G_2}=1-w_{D_1}}. \end{array} \end{equation}

Otherwise, if \eqref{501} results in $\sigma _{G_2}^{2}=\sigma _{f}^{2}(d_{G_2})$ with $d_{G_2}<d_{max}$ (case $\mathcal{K}$=3 in Fig. \ref{fig:pD2_BG_Fig1}), the variance and scaling factor of the second Gaussian component are expressed by
\begin{equation} \label{5022} \sigma _{G_2}^{2}=\sigma _{f}^{2}(d_{G_2}), \quad w_{G_2}=\frac{f_{D_{I} }(d=0)}{f_{G_{2} }(d=0)}. \end{equation}

\noindent
- step 3.

Given $\sigma _{G_2}^{2}$ and $w_{G_2}$ in \eqref{5022}, the residual distribution $f_{D_{I_{2}}}(d)$ is given by
\begin{equation} \label{503} f_{D_{I_{2}}}(d)= f_{D_{I_1}}(d)-w_{G_2} f_{G_{2}}(d). \end{equation}

Given the residual distribution $f_{D_{I_{2}}}(d)$ \eqref{503}, we compute the point $d_{G_3}$ where $f_{D_{I_{2}}}(d)$ falls to $c_{0} f_{D_{I}}(d)$, in order to estimate where an additional Gaussian component would be beneficial to the distribution. Thus, we obtain
\begin{equation} \label{504} d_{G_3}=\mathop{\arg }\limits_{d} \left\{f_{D_{I_2} } (d)-c_{0} f_{D_I} (d)\right\}. \end{equation}

\begin{figure}[h]
  \centering
    \includegraphics[width=9.0cm, height=6.8cm]{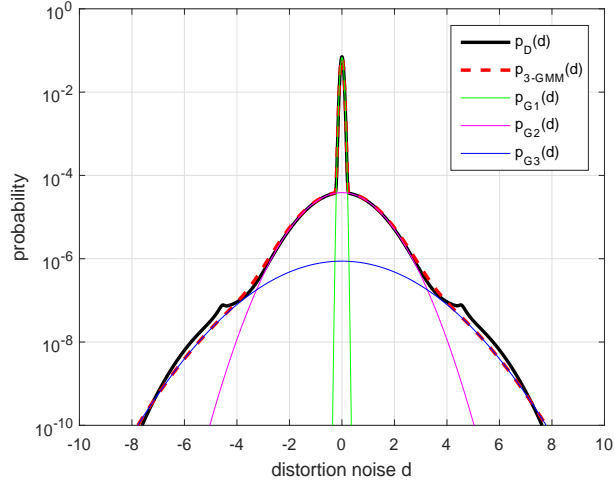}
  \caption{Example of $f_{D}(d)$ and its approximation by 3-GMM, at SIR=-30dB, SNR=25dB and $p_I=0.001$, for OFDM with 256 subcarriers and 4QAM affected by S$\alpha$S ImpN at $\alpha=1.2$.}
  \label{fig:pD4_SaS_Fig1}
\end{figure}

Given $d_{G_3}$ in \eqref{504}, we can compute the variance and scaling factor of the third Gaussian component by\footnote{slightly improved estimate of $\sigma _{G_3}^{2}$ is obtained by averaging with $\sum _{i=i_{3+1}}^{2 i_{3}}\sigma _{f}^{2}(d_{i})$, where $i_{3}=\mathop{\arg }\limits_{i} \left\{d_{G_3}\right\}.$}
\begin{equation}  \label{505} \sigma _{G_3}^{2}=\sigma _{f}^{2}(d_{G_3}), \quad w_{G_3}=\frac{f_{D_{I} }(d=d_{G_3})}{f_{G_{2} }(d=d_{G_3})}. \end{equation}

Given \eqref{502} and \eqref{505}, $f_{D_{I} } (d)$ can be approximated by a 2-GMM, and thus the total $f_{D} (d)$ can be well approximated by $\mathcal{K}=3$, i.e., a 3-GMM (Fig. \ref{fig:pD4_SaS_Fig1}) with the following mixture parameters 
\begin{equation} \label{506} \begin{array}{l} {\sigma _{G_1}^{2}=\sigma _{D_1}^{2}, \quad \sigma _{G_2}^{2}=\sigma _{f}^{2}(d_{G_2}),\quad \sigma _{G_3}^{2}=\sigma _{f}^{2}(d_{G_3})}, \\ {w_{G_1}=w_{D_1}, \quad w_{G_2}=\frac{f_{D_{I} }(d=0)}{f_{G_{2} }(d=0)}, \quad w_{G_3}=\frac{f_{D_{I} }(d=d_{G_3})}{f_{G_{2} }(d=d_{G_3})} }. \end{array} \end{equation}

\noindent
- step 4.

Given $\sigma _{G_3}^{2}$ and $w_{G_3}$, the residual distribution is obtained by
\begin{equation} \label{507} f_{D_{I_{3}}}(d)= f_{D_{I_{2}}}(d)-w_{G_3} f_{G_{3}}(d). \end{equation}

Given the residual distribution $f_{D_{I_{3}}}(d)$ \eqref{507}, we compute the point $d_{G_4}$ where $f_{D_{I_{3}}}(d)$ falls to $c_{0} f_{D_{I}}(d)$ in order to estimate where an additional Gaussian component would be beneficial to the distribution approximation. Thus, we compute
\begin{equation} \label{508} d_{G_4}=\mathop{\arg }\limits_{d} \left\{f_{D_{I_3} } (d)-c_{0} f_{D_I} (d)\right\} \end{equation}

\noindent
and, if $d_{G_4}=d_{max}$, we conclude that an additional Gaussian component is not needed and the GMM model is completed by \eqref{506}, approximating $f_{D}(d)$ by a 3-GMM.

Otherwise, if $d_{G_4}>d_{max}$, the additional Gaussian component is considered (case $\mathcal{K}=4$ in Fig. \ref{fig:pD2_BG_Fig1}), with a variance and scaling factor expressed by
\begin{equation}  \label{509} \sigma _{G_4}^{2}=\sigma _{f}^{2}(d_{G_4}), \quad w_{G_4}=\frac{f_{D_{I} }(d=d_{G_4})}{f_{G_{2} }(d=d_{G_4})}. \end{equation}

Finally, given \eqref{506} and \eqref{509}, the total $f_{D} (d)$ is  approximated by $\mathcal{K}=4$, i.e. 4-GMM with mixture parameters expressed by 
\begin{equation} \label{510)} \begin{array}{l} {\sigma _{G_1}^{2}=\sigma _{D_1}^{2}, \sigma _{G_2}^{2}=\sigma _{f}^{2}(d_2), \sigma _{G_3}^{2}=\sigma _{f}^{2}(d_{G_3}), \sigma _{G_4}^{2}=\sigma _{f}^{2}(d_{G_4})} \\ {w_{G_1}=w_{D_1}, \quad w_{G_2}=\frac{f_{D_{I} }(d=0)}{f_{G_{2} }(d=0)}, \quad w_{G_3}=\frac{f_{D_{I} }(d=d_{G_3})}{f_{G_{2} }(d=d_{G_3})}}, \\ {w_{G_4}=\frac{f_{D_{I} }(d=d_{G_4})}{f_{G_{2} }(d=d_{G_4})}}. \end{array} \end{equation}

Evidently, expressions \eqref{500}, \eqref{502}, \eqref{506}, and \eqref{510)}, are the closed-form formulas to compute the variances and the weighting factors for $\mathcal{K} \in \{1, 2, 3, 4 \}$ in \eqref{50)}, to approximate the distortion noise distribution $f_{D}(d)$ by a proper $\mathcal{K}$-GMM.

\label{app:appendixC}

\section{An Example of the Closed-form Approximation of a Given 4-GMM $pdf$ } 

This Appendix illustrates an example of the closed-form approximation of a 4-GMM $pdf$ $f_{Z}(z)$, with $\sigma_k^2 \in \left\{0.0032,\, 10,\, 100,\,1000\right\}$ and $p_k \in \left\{0.90009,\,0.09,\, 0.0099,\, 0.00001\right\}$, by $\mathcal{K}$-GMM which  leverages on a component-by-component Gaussian fitting, as detailed in Appendix B. Assuming that $f_{Z}(z)$ is belongs to the family of Gaussian $pdf$s, the variance distribution can be calculated from two neighbour samples $z_i$ and $z_{i+1}$ of $\left\{f_{Z}(z_i)\right\}_{i=1,\cdots,M_{z}}$ 
by
\begin{equation} \label{ZEqnNum415000} \sigma _{f}^{2}(z_i) =0.5\frac{z_{i+1}^{2} -z_{i}^{2} }{\log f_{Z} (z_{i} )-\log f_{Z} (z_{i+1} )}.  \end{equation} 
\noindent

\begin{figure}[h]
  \centering
  \includegraphics[width=8.0cm, height=6.0cm]{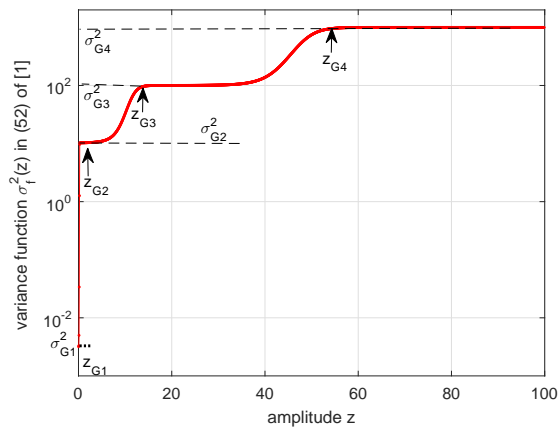}
  \caption{Variance function $\sigma _{f}^{2}(z_i)$ obtained by \eqref{ZEqnNum415000} for the 4-GMM example, as the support vector used to estimate the Gaussian components.} 
  \label{fig:varf_Fig1}
\end{figure}

Fig. \ref{fig:varf_Fig1} shows the right-side part of the variance function computed for $M_z$ points, thus $\left\{\sigma _{f}^{2}(z_i)\right\}_{i=1,\cdots,M_{z}}$, which we exploit to estimate the Gaussian components i.e., their variances $\sigma _{G_k}^{2} $ and scaling weights $w _{G_k}$ for $k=1, 2, 3, 4$.
The left-side of the variance function $\sigma _{f}^{2}(z_i)$ can be also computed by \eqref{ZEqnNum415000}, however it is expectedly symmetric one, likewise to the itself $f_{Z}(z)$.

\begin{figure}[h]
  \centering
    \includegraphics[width=8.0cm, height=6.0cm]{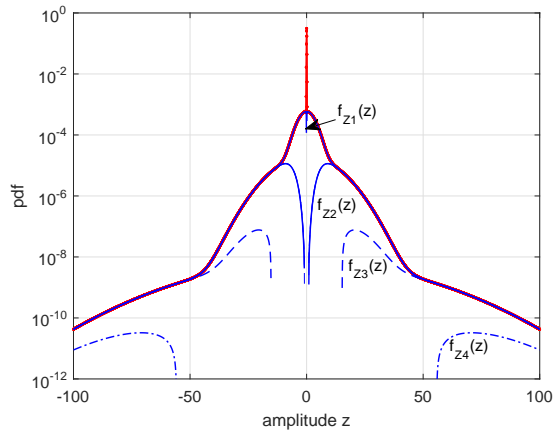}
  \caption{Plot of the components used to approximate of $f_{Z}(z)$ by a $\mathcal{K}$-GMM, by means of (55),(56); (60),(61); (64),(65), as detailed in Appendix B.}
  \label{fig:pDk_Fig2}
\end{figure}

Fig. \ref{fig:pDk_Fig2} shows the intermediate results obtained by means of (56), (60) and (64), based on which are computed variances and scaling factors by means of (54), (58), (63) and (67).

\begin{figure}[h]
  \centering
    \includegraphics[width=8.0cm, height=6.0cm]{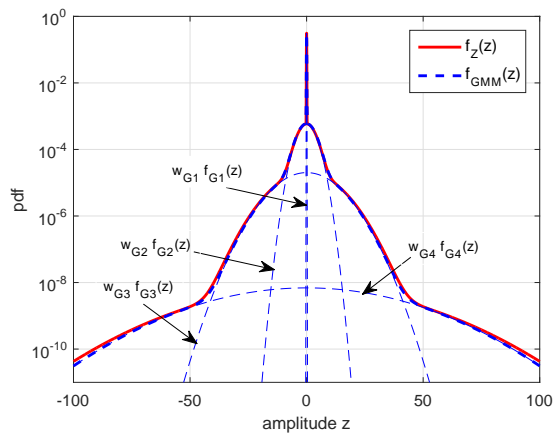}    
  \caption{Comparison of $f_{Z}(z)$ and its 4-GMM approximation obtained by steps 1 to 4 presented in Appendix B.}
  \label{fig:pGk_Fig3}
\end{figure}

Fig. \ref{fig:pGk_Fig3} compares the original 4-GMM and the fitted 4-GMM $pdf$s, also showing the four estimated weighted Gaussian components $w_{G_k} f_{G_k}(z)$ for $k=1, 2, 3, 4$.

The results in Fig. \ref{fig:pGk_Fig3} show that the given $pdf$ can be well approximated by $\mathcal{K}$-GMM built following the closed-form algorithm presented in Appendix B. Note, that the better fitting results can be expected only by performing an iterative optimization algorithm, which however is out of the scope of this work. Actually, GMM approximation of the distortion noise $pdf$ at the output of the NLS, considered in this work, is used to obtain the GMM-based closed-form formula for SER prediction.

\bibliographystyle{IEEEtran}

\bibliography{./bib/reference_SER}

\begin{thebibliography}{10}
\providecommand{\url}[1]{#1}
\csname url@samestyle\endcsname
\providecommand{\newblock}{\relax}
\providecommand{\bibinfo}[2]{#2}
\providecommand{\BIBentrySTDinterwordspacing}{\spaceskip=0pt\relax}
\providecommand{\BIBentryALTinterwordstretchfactor}{4}
\providecommand{\BIBentryALTinterwordspacing}{\spaceskip=\fontdimen2\font plus
\BIBentryALTinterwordstretchfactor\fontdimen3\font minus
  \fontdimen4\font\relax}
\providecommand{\BIBforeignlanguage}[2]{{%
\expandafter\ifx\csname l@#1\endcsname\relax
\typeout{** WARNING: IEEEtran.bst: No hyphenation pattern has been}%
\typeout{** loaded for the language `#1'. Using the pattern for}%
\typeout{** the default language instead.}%
\else
\language=\csname l@#1\endcsname
\fi
#2}}
\providecommand{\BIBdecl}{\relax}
\BIBdecl

\bibitem{Ghosh_1996}
M.~Ghosh, ``Analysis of the effect of impulse noise on multicarrier and single
  carrier {QAM} systems,'' \emph{IEEE Trans. Commun.}, vol.~44, no.~2, pp.
  145--147, 1996.

\bibitem{Price_1958}
R.~Price, ``A useful theorem for nonlinear devices having {G}aussian inputs,''
  \emph{IEEE Trans. Inf. Theory}, vol.~4, no.~8, pp. 69--72, 1958.

\bibitem{Minkoff_1985}
J.~Minkoff, ``The role of {AM}-to-{PM} conversion in memoryless nonlinear
  systems,'' \emph{IEEE Trans. on Communications}, vol.~33, no.~2, pp.
  139--144, 1985.

\bibitem{Banelli_2000}
P.~Banelli and S.~Capobardi, ``Theoretical analysis of performance of {OFDM}
  signals in non-linear {AWGN} channels,'' \emph{IEEE Trans. on
  Communications}, vol.~48, no.~3, pp. 430--441, 2000.

\bibitem{Miyamoto_1995}
S.~Miyamoto, M.~Katayama, and N.~Morinaga, ``Performance analysis of {QAM}
  systems under {C}lass-{A} noise environment,'' \emph{IEEE Trans. on
  Electromag. Comp.}, vol.~37, no.~2, pp. 260--267, 1995.

\bibitem{Ma_2005}
Y.~H. Ma, P.~So, and E.~Gunawan, ``Performance analysis of {OFDM} systems for
  broadband power line communications under impulsive noise and multipath
  effects,'' \emph{IEEE Trans. Power Delivery}, vol.~20, no.~2, pp. 674--682,
  2005.

\bibitem{Amirshahi_2006}
P.~Amirshahi, S.~M. Navidpour, and M.~Kavehrad, ``Performance analysis of
  uncoded and coded {OFDM} broadband transmission over low voltage power-line
  channels with impulsive noise,'' \emph{IEEE Trans. on Power Delivering},
  vol.~21, no.~4, pp. 1927--1934, 2006.

\bibitem{Middleton_1977}
D.~Middleton, ``Statistical-physical models for electromagnetic interference,''
  \emph{IEEE Trans. Electromagn. Compat.}, vol.~19, no.~4, pp. 106--127, 1977.

\bibitem{Middleton_1999}
------, ``Non-gaussian noise models in signal processing for
  telecommunications: New methods and results for {C}lass-{A} and {C}lass-{B}
  noise models,'' \emph{IEEE Trans. Inf. Theory}, vol.~45, no.~4, pp.
  1129--1149, 1999.

\bibitem{Shao_1993}
M.~Shao and C.~Nikias, ``Signal processing with fractional lower order moments:
  stable processes and their applications,'' \emph{Proceedings of the IEEE},
  vol.~81, no.~7, pp. 986 --1010, July 1993.

\bibitem{Seo_1989}
J.~S. Seo, S.~J. Cho, and K.~Feher, ``Impact of non-{Gaussian} impulsive noise
  on the performance of high-level {QAM},'' \emph{IEEE Transactions Electromag.
  Compatibility}, vol.~31, no.~2, pp. 177--180, 1989.

\bibitem{Haring_2004}
J.~H\"{a}ring and A.~J.~H. Vinck, ``Coding and signal space diversity for a
  class of fading and impulsive noise channels,'' \emph{IEEE Trans. Inform.
  Theory}, vol.~50, no.~5, pp. 887--895, 2004.

\bibitem{Zhidkov_2006}
S.~Zhidkov, ``Performance analysis and optimization of {OFDM} receiver with
  blanking nonlinearity in impulsive noise environment,,'' \emph{IEEE Trans.
  Veh. Techn.}, vol.~55, no.~1, pp. 234--242, Jan. 2006.

\bibitem{Zhidkov_2008}
------, ``Analysis and comparison of several simple impulsive noise mitigation
  schemes for {OFDM} receivers,'' \emph{IEEE Trans. Commun.}, vol.~56, no.~1,
  pp. 5--9, 2008.

\bibitem{Banelli_2015}
P.~Banelli and L.~Rugini, ``Impulsive noise mitigation for wireless {OFDM},''
  in \emph{Proc. IEEE Int. Workshop on Sig. Proc. Adv. in Wireless Comm.,
  (SPAWC)}, Stockholm, Sweden, June-July 2015, pp. 346--350.

\bibitem{Darsena_2015}
D.~Darsena, G.~Gelli, F.~Melito, and F.~Verde, ``{ICI}-free equalization in
  {OFDM} systems with blanking preprocessing at the receiver for impulsive
  noise mitigation,'' \emph{IEEE Signal Proc. Lett.}, vol.~22, no.~9, p. 1321,
  2015.

\bibitem{Banelli_2016}
L.~Rugini and P.~Banelli, ``On the equivalence of maximum {SNR} and {MMSE}
  estimation: Applications to additive non-{G}aussian channels and quantized
  observations,'' \emph{IEEE Trans. on Signal Processing}, vol.~64, no.~23, pp.
  6190--6199, 2016.

\bibitem{Rozic_2017}
N.~Ro\v{z}i\'{c}, P.~Banelli, D.~Begu$\check{\rm{s}}$i\'{c}, and J.~Radi\'{c},
  ``Multiple threshold-based estimators for impulsive noise suppression in
  multicarrier communications,'' \emph{IEEE Trans. on Sign. Processing},
  vol.~66, no.~6, pp. 1619--1633, 2018.

\bibitem{Banelli_2013}
P.~Banelli, ``Bayesian estimation of a gaussian source in {M}iddleton
  {C}lass-{A} impulsive noise,'' \emph{IEEE Signal Proc. Lett.}, vol.~20,
  no.~10, pp. 956--959, 2013.

\bibitem{Vastola_1984}
K.~S. Vastola, ``Threshold detection in narrow-band non-{G}aussian noise,''
  \emph{IEEE Trans.Commun.}, vol.~32, no.~2, pp. 134--139, Feb. 1984.

\bibitem{Suraweera_2003}
H.~A. Suraweera, C.~Chai, J.~Shentu, and J.~Armstrong, ``Analysis of impulse
  noise mitigation techniques for digital television systems,'' in \emph{in
  Proc. 8th Int. OFDM Workshop}, Lisbon, Portugal, Sept. 2003 2003, pp.
  172--176.

\bibitem{Gao_2007}
P.~Gao and C.~Tepedelenlioglu, ``Space-time coding over fading channels with
  impulsive noise,'' \emph{IEEE Trans. Wireless Commun.}, vol.~6, no.~1, pp.
  220--229, 2011.

\bibitem{Savoia_2013}
R.~Savoia and F.~Verde, ``Performance analysis of distributed space-time block
  coding schemes in {M}iddleton {C}lass-{A} noise,'' \emph{IEEE Trans. Veh.
  Techn.}, vol.~62, no.~6, pp. 2579--2595, 2013.

\bibitem{Rugini_2005}
L.~Rugini and P.~Banelli, ``{BER} of {OFDM} systems impaired by carrier
  frequency offset in multipath fading channels,'' \emph{IEEE Trans. on
  Wireless Comm.}, vol.~4, no.~5, pp. 2279--2288, 2005.

\bibitem{Eriksson_1995}
H.~B. Eriksson, P.~Odling, T.~Koski, and P.~Borjesson, ``A genue-aided detector
  with a probabilistic description of the side information,'' in \emph{in Proc.
  of the IEEE International Symposium on Information Theory}, Whistler, BC,
  Canada, Canada, 17-22 Sept. 1995, pp. 332--332.

\bibitem{Wang_2000}
Z.~Wang and G.~B. Giannakis, ``Wireless multicarrier communications,''
  \emph{IEEE {S}ignal {P}rocessing {M}agazine}, vol.~17, no.~3, pp. 29--48,
  2000.

\bibitem{Banelli_2014}
P.~Banelli and L.~Rugini, ``{OFDM} and multicarrier signal processing,'' in
  \emph{Academic Press Library in Signal Processing: Vol.2 - Communications and
  Radar Signal Processing}.\hskip 1em plus 0.5em minus 0.4em\relax Elsevier,
  2014, vol.~2, pp. 187--293.

\bibitem{Rice_1948}
S.~O. Rice, ``Statistical properties of a sine wave plus random noise,''
  \emph{Bell Syst. Techn. J.}, vol.~27, pp. 109--157, 1948.

\bibitem{Nakagami_1960}
M.~Nakagami, \emph{The m-distribution, a general formula of intensity
  distribution of rapid fading}.\hskip 1em plus 0.5em minus 0.4em\relax in
  Statistical Methods in Radio Wave Propagation, New York: Pergamon, 1960,
  vol.~2.

\bibitem{Kassam_1988}
S.~A. Kassam, \emph{Signal Detection in Non-{G}aussian Noise}, 1st~ed.\hskip
  1em plus 0.5em minus 0.4em\relax Philadelphia, USA: New York, Berlin, Dowden
  and Culver, Inc, 1988.

\bibitem{Caire_2014}
T.~Y. Al-Naffouri, A.~A. Quadeer, and G.~Caire, ``Impulse noise estimation and
  removal for {OFDM} systems,'' \emph{IEEE Trans. on Communications}, vol.~62,
  no.~3, pp. 976--989, 2014.

\bibitem{Hur_2016}
S.~Hur, S.~Baek, B.~Kim, Y.~Chang, A.~F. Molisch, T.~S. Rappaport, K.~Haneda,
  and J.~Park, ``Proposal on millimeter-wave channel modeling for 5{G} cellular
  system,'' \emph{IEEE J. of Select. Topics in Signal Process.}, vol.~10,
  no.~3, pp. 454--469, March 2016.

\bibitem{Blackard_1993}
K.~L. Blackard, T.~S. Rappaport, and C.~W. Bostian, ``Measurement and models of
  radio frequency impulsive noise for indoor wireless communications,''
  \emph{IEEE J. Sel. Areas Commun.}, vol.~11, no.~9, pp. 991--1001, Sept. 1993.

\bibitem{Sanchez_1999}
M.~G. Sanchez and et~al., ``Impulsive noise measurements and characterization
  in a {UHF} digital {TV} channel,'' \emph{IEEE Trans. Electromagn. Compat.},
  vol.~41, no.~2, pp. 124--136, May 1999.

\bibitem{Lin_2013}
M.~N. J.Lin and B.~L. Evans, ``Impulsive noise mitigation in powerline
  communications using sparse {B}ayesian learning,'' \emph{IEEE J. Sel. Areas
  Commun.}, vol.~31, no.~7, pp. 1172--1183, July 2013.

\bibitem{Feller_1971}
W.~Feller, \emph{An Introduction to Probability Theory and its Application,
  Vol. II}.\hskip 1em plus 0.5em minus 0.4em\relax New York: John Wiley \&
  Sons, Inc., 1971.

\bibitem{Hamdan_2004}
H.~N. Hamdan and J.~P. Nolan, ``Estimating the parameters of infinite scale
  mixtures of normals,,'' in \emph{Proceedings of the 36th Symposium on the
  Interface: Computing Science and Statistics}, 2004.

\bibitem{Kuruoglu_1998}
E.~E. Kuruoglu, C.~Molina, and W.~J. Fitzgerald, ``Approximation of
  alpha-stable probability densities using finite {G}aussian mixtures,'' in
  \emph{Proc. EUSIPCO 1998}, Rhodes, Greece, Sept. 1998.

\bibitem{Rozic_2013b}
N.~Rozic, M.~Russo, M.~Saric, J.~Radic, and D.~Begusic, ``Short report on
  approximation of the symmetric alpha-stable density by {GMM},
  p023-02311924-060212,
  http://marjan.fesb.hr/~radic/technicalreports/sas.pdf,'' University of Split,
  FESB, Tech. Rep., 2013.

\bibitem{Evans_2000}
M.~Evans, N.~Hastings, and B.~Peacock, \emph{Statistical Distributions},
  3rd~ed.\hskip 1em plus 0.5em minus 0.4em\relax Wiley, 2000.

\bibitem{Simon_2006}
M.~K. Simon, \emph{Probability Distributions Involving {G}aussian Random
  Variables: A Handbook for Engineers, Scientists and Mathematicians}.\hskip
  1em plus 0.5em minus 0.4em\relax Secaucus, NJ, USA: Springer-Verlag New York,
  Inc., 2006.

\bibitem{Proakis_2008}
J.~Proakis, \emph{Digital Communications}, 5th~ed.\hskip 1em plus 0.5em minus
  0.4em\relax New York, NY, USA: New York: Mc. Graw-Hill, 2008.

\bibitem{Bussgang_1952}
J.~J. Bussgang, \emph{Cross correlation function of amplitude-distorted
  Gaussian signals, Tech. Rep. 216}.\hskip 1em plus 0.5em minus 0.4em\relax
  M.I.T Cambridge, March 1952.

\bibitem{Alouini_1999}
M.-S. Alouini and A.~Goldsmith, ``A unified approach for calculating error
  rates of linearly modulated signals over generalized fading channels,''
  \emph{IEEE Trans. Commun.}, vol.~47, no.~92, pp. 1324--1334, 1999.

\bibitem{Yen_2007}
R.~Y. Yen, H.-Y. Liu, K.-F. Yang, M.-C. Hong, and S.-C. Nan, ``Performance of
  {OFDM} {QAM} over frequency-selective fading channels,'' in \emph{in Proc.
  IEEE 2007 IFIP International Conference on Wireless and Optical
  Communications Networks (WOCN 2007.)}, Singapore, Singaporel, 2-4 July 2007
  2007, pp. 1--5.

\bibitem{Craig_1991}
J.~W. Craig, ``A new, simple, and exact result for calculating the probability
  of error for two-dimensional signal constellations,'' in \emph{in Proc. of
  the IEEE Military Communications Conf. (MILCOM�91)}, vol.~1, McLean, VA,,
  october 1991, pp. 571--575.

\bibitem{Shayesteh_1995}
M.~G. Shayesteh and A.~Aghamohammadi, ``On the error probability of linearly
  modulated signals on frequency-flat {R}icean, {R}ayleigh, and {AWGN}
  channels,'' \emph{IEEE Trans. Commun.}, vol.~43, no. 2/3/4, pp. 1454--1466,
  19959.

\bibitem{Hughes_2005}
D.~Hughes-Hallet and W.~G. McCullum, \emph{Calculus}, 4th~ed.\hskip 1em plus
  0.5em minus 0.4em\relax Wiley, 2005.

\end{thebibliography}

\end{document}